\documentclass[12pt]{article}
\usepackage{comment}
\usepackage{latexsym,makeidx,graphicx,subfig,rotate}
\usepackage[centertags]{amsmath}
\usepackage[active]{srcltx}
\usepackage{amsfonts}
\usepackage{amssymb}
\usepackage{amsthm}
\usepackage{newlfont}
\usepackage{amsfonts}
\usepackage{amssymb}
\usepackage{indentfirst}
\usepackage{graphics}
\usepackage{amsmath}
\usepackage{natbib}
\allowdisplaybreaks

\newfont{\Sc}{eusm10}

\newcommand{\cip}{\mbox{$\perp\!\!\!\perp$}}
\newcommand{\cb}{ \begin{eqnarray} }
\newcommand{\ce}{ \end{eqnarray} }
\newcommand{\expit}{ \mbox{\rm expit} }
\newcommand{\logit}{ \mbox{\rm logit} }
\newcommand{\Bern}{ \mbox{\rm Bern} }

\begin{document}

\title{Assumption-lean inference \\ for generalised linear model parameters}
\author{Stijn Vansteelandt$^{1,2}$ and Oliver Dukes$^1$\\
$^1$: Ghent University,
Ghent,
Belgium\\
$^2$: London School of Hygiene and Tropical Medicine, London, U.K.}
\date{\today}

\maketitle

Inference for the parameters indexing generalised linear models is routinely based on the assumption that the model is correct and a priori specified. 
This is unsatisfactory because the chosen model is usually the result of a data-adaptive model selection process, which may induce excess uncertainty that is not usually acknowledged. Moreover, the assumptions encoded in the chosen model rarely represent some a priori known, ground truth, making standard inferences prone to bias, but also failing to give a pure reflection of the information that is contained in the data. Inspired by developments on assumption-free inference for so-called projection parameters, we here propose novel nonparametric definitions of main effect estimands and effect modification estimands. These reduce to standard main effect and effect modification parameters in generalised linear models when these models are correctly specified, but have the advantage that they continue to capture respectively the primary (conditional) association between two variables, or the degree to which two variables interact (in a statistical sense) in their effect on outcome, even when these models are misspecified. We achieve an assumption-lean inference for these estimands (and thus for the underlying regression parameters) by deriving their influence curve under the nonparametric model and invoking flexible data-adaptive (e.g., machine learning) procedures. \\

Key words: bias; conditional treatment effect; estimand; influence curve; interaction; model misspecification; nonparametric inference.

\section{Introduction}

Statistical analyses routinely invoke modelling assumptions. These include smoothness assumptions, implied by parametric or semi-parametric model specifications, for instance, but also sparsity assumptions that underlie variable selection procedures. Such assumptions are generally a necessity.
The curse of dimensionality indeed forces one to borrow information across strata of subjects with different covariate values, as well as to reduce the dimensions of the possibly many measured variables. Modelling assumptions are often also a deliberate choice. With a continuous exposure, for instance, one would often not be interested in knowing exactly how the outcome changes with each increase in exposure, but might content oneself with a `simple' and parsimonious summary of the exposure effect. Models enable one to create such summaries. This distinction in the nature of the assumptions is rarely made in how we approach a data analysis, but is nonetheless an essential one that will turn out key to the strategy that we advocate.

Regardless of this distinction, modelling assumptions are almost always a pure mathematical convenience, and not reflecting a priori knowledge that we had prior to seeing the data. Ideally,  in such cases, their analysis should therefore only extract information from the data, and not from the assumptions.
This realisation is not new. It  
became very dominant in the 90's in work on non-ignorable incomplete data.
Rotnitzky and Robins (e.g., \cite{rotnitzky1997analysis,rotnitzky1998semiparametric,scharfstein1999adjusting}), amongst others, then increased awareness that modelling assumptions, such as normality and linearity assumptions,
may sometimes permit identification of parameters in the absence of missing data assumptions.
There is now a fairly general agreement that such identification is dishonest 
when these modelling assumptions are made for convenience.
In spite of this, once we have stated those structural assumptions needed for identification,  
we often fall back into our routine. We continue to rely on modelling assumptions more than we may realise, and treat them as representing some ground truth in how we approach inference.

For instance, likelihood-based or semi-parametric estimation approaches 
extract information not only from the data, but also from the modelling assumptions as if they were given. In fact, maximum likelihood estimators, maximum a posteriori estimators and semi-parametric efficient estimators precisely succeed to increase efficiency by taking modelling assumptions as given, and extracting information from them.
It makes the resulting data analysis no longer purely evidence-based.
We usually try to make up for that by adopting model or variable selection procedures. However, 
the inferences that are commonly provided, continue to pretend that the model delivered by these procedures, 
was a priori given and known, which can sometimes make things worse. All of this is raising questions over the `honesty' of the data analyses that we produce.

Motivated by these concerns, enormous progress has been made over the past several decades in terms of how to develop an inference that is `assumption-free', across several different literatures. \citet{white_heteroskedasticity-consistent_1980} developed the so-called `sandwich estimator' of the standard error for ordinary least squares (OLS); this delivers a valid measure of uncertainty around the regression coefficient estimates, even if the model-based assumptions of OLS (linearity, heteroscedasticity) are not met. \citet{freedman_so-called_2006} noted that although the sandwich estimator is unbiased under nonlinearity, the resulting confidence intervals and tests are not useful given that it may be unclear what the model coefficients represent. Several proposals for restoring meaning to regression estimates have been made, seeing a model coefficient as a projection parameter \citep{buja_models_2019}, or variable importance measure \citep{chambaz_estimation_2012}, both ideas which have gained traction in high-dimensional statistics \citep{berk_valid_2013,wasserman_discussion_2014}. In terms of doing causal inference, \citet{lin_agnostic_2013} gave a `model-agnostic' approach to the adjustment for baseline covariates in randomised experiments. He noted that ``one does not need to believe in the classical linear model to tolerate or even advocate OLS adjustment." Related work has explored how OLS estimates can in certain settings be interpreted as weighted averages of treatment effects, even when the linear model is wrong  \citep{angrist_empirical_1999,angrist_mostly_2009,aronow_does_2016,graham_semiparametrically_2018,sloczynski2020interpreting}. Many of the above approaches start with a common estimator of a parameter indexing a parametric regression model. They then characterise what estimand corresponds to the limit of the estimator, regardless of whether the model is true. In contrast, Mark van der Laan and collaborators take an alternative approach in their scientific `roadmap' \citep{van_der_laan_targeted_2006,van_der_laan_targeted_2011}. They first define an estimand as a functional of the observed data distribution, which characterises what we aim to infer from the data, and next develop estimation and inference based on the inference curve (provided the estimand is pathwise-differentiable under the nonparametric model; see Section 5), with all nuisance functionals estimated non-parametrically (e.g., via machine learning). The use of influence curves is essential to this development, as it enables valid inference even when the analysis is based on data-adaptive procedures, such as machine learning, variable selection, model selection, etc. Attention is mainly given to causal inference applications where the choice of summary estimand may be relatively straightforward e.g. the average causal effect of a binary treatment on a clinical outcome. 

Key to the latter developments is to change the starting point of the analysis from the postulation of a statistical model to the postulation of an estimand. This change of focus brings many advantages. It forces one to work with well-understood estimands from the start. It enables one to separate modelling assumptions made for parsimony, which will be used to define the estimand, from assumptions imposed to handle the curse of dimensionality. It prevents reliance on these assumptions, as inference for the estimand can be developed under the nonparametric model. Finally, the resulting analysis can be pre-specified, which is essential if one aims for an honest data analysis that reflects all uncertainties, including the uncertainty surrounding the model that is used. 

Changing this focus of the analysis is non-trivial, however. It turns the difficulty of postulating a model, to which we have grown to become familiar,
into the difficulty of choosing an estimand, for which infinitely many choices can typically be conceived. While there is some experience in choosing meaningful estimands in causal inference applications, complications easily arise when e.g. considering continuous exposures, or when general association measures (e.g. measures of a time trend) rather than causal effect measures are of interest. It calls for the development of specific estimands that can be used quite generically (in a sense that we will make specific later) and connect to regression parameters that practitioners have grown to become familiar with. In this way, they can provide an assumption-lean inference for those standard regression parameters, which uses the underlying model only with the aim to summarise and deliver a familiar interpretation, but relates to machine learning procedures running in the background to assure valid inference. 
In this paper, we will show how this is ideally done when the aim is to infer regression parameters indexing generalised linear models. In particular, we propose novel estimands for conditional association measures between two variables, and for the degree to which two variables interact (in a statistical sense) in their effect on outcome, which are well defined in a nonparametric sense (i.e., regardless of what is the underlying data-generating distribution). We achieve an assumption-lean inference for these estimands by deriving their influence curve under the nonparametric model and invoking flexible data-adaptive (e.g., machine learning) procedures. Since the proposed estimands reduce to standard main effect and interaction parameters in arbitrary generalised linear models when these models are correctly specified, we thus generalise standard inference to give a pure reflection of the information that is contained in the data. Our developments thus provide a novel framework for fitting generalised linear models, and at a broader level, also shed light on what defines an adequate estimand, and how it can be constructed.

\section{Illustration}\label{sec:ill}

To clarify the points made in the introduction, we provide a simple illustration with artificial, independent data for $n=50$ subjects on a scalar standard normal variate $L$, a dichotomous exposure $A$, coded 0 or 1, with $P(A=1|L)=\mbox{\rm expit}(L-L^2)$ and a normally distributed outcome with mean $A-L+4.5AL+0.5L^2-2.25AL^2$ and unit (residual) variance. 
The ordinary least squares estimator for $\beta$ under model 
\[E(Y|A,L)=\alpha_0+\alpha_1L+\beta A,\]
can be shown to converge to 
\begin{eqnarray*}
&&\frac{E\left[\pi(L)\left\{1-\tilde{\pi}(L)\right\}\left\{E(Y|A=1,L)-E(Y|A=0,L)\right\}\right]}{E\left[\pi(L)\left\{1-\tilde{\pi}(L)\right\}\right]}\\&&\hspace*{1cm}+\frac{E\left[\left\{\pi(L)-\tilde{\pi}(L)\right\}E(Y|A=0,L)\right]}{E\left[\pi(L)\left\{1-\tilde{\pi}(L)\right\}\right]},\end{eqnarray*}
where $\pi(L)=P(A=1|L)$ and $\tilde{\pi}(L)$ denotes the population least squares projection of $A$ onto 1 and $L$. This displayed `estimand' consists of two contributions. The first is a weighted average of the contrasts $E(Y|A=1,L)-E(Y|A=0,L)$. It is informative about the conditional association between $A$ and $Y$. The second contribution is a weighted average of the contrasts $\pi(L)-\tilde{\pi}(L)$. It is not informative about the conditional association between $A$ and $Y$ and is generally non-zero, except when the linear outcome model is correctly specified or $\pi(L)$ happens to be a linear function of $L$ (see e.g. \cite{robins1992estimating,vansteelandt2014structural}). This is disturbing. It makes the estimand targeted by the ordinary least squares estimator a questionable summary of the conditional association between $A$ and $Y$, given $L$, when the linear model is misspecified. 

A more attractive approach is based on the partially linear model
\begin{equation}\label{linear}
E(Y|A,L)=\omega(L)+\beta A,\end{equation}
where $\beta$ and $\omega(L)$ are unknown. Here, $\hat{\beta}$ can be obtained as the E-estimator 
\begin{align}\label{eestimator}
\frac{\sum_{i=1}^n \left\{A_i-\hat{\pi}(L_i)\right\}\left\{Y_i-\hat{\omega}(L_i)\right\}}{\sum_{i=1}^n \left\{A_i-\hat{\pi}(L_i)\right\}A_i},
\end{align}
\citep{robins1992estimating}, where $\hat{\pi}(.)$ and $\hat{\omega}(.)$ are possibly data-adaptive (e.g. machine learning-based) estimators of $\pi(.)$ and $\omega(.)$, respectively. In the illustration in the next paragraph, for instance, we have based $\pi(.)$ on a logistic additive model and estimated $\omega(.)$ using smoothing splines in model (\ref{linear}). The ability to use data-adaptive procedures, makes it more plausible to reason under the assumption that $\hat{\pi}(.)$ converges to $\pi(.)$, which we will make. In that case, the above estimator has been shown \citep{vansteelandt2014regression} to converge to the weighted contrast
\begin{equation}\label{eq:estimand}
\frac{E\left[\pi(L)\left\{1-{\pi}(L)\right\}\left\{E(Y|A=1,L)-E(Y|A=0,L)\right\}\right]}{E\left[\pi(L)\left\{1-{\pi}(L)\right\}\right]},\end{equation}
of the conditional outcome mean at $A=1$ versus $A=0$, 
even when model (\ref{linear}) is misspecified, e.g. because $A$ and $L$ interact in their effect on outcome.  

It follows from the above reasoning that the E-estimator, as opposed to the ordinary least squares estimator, is not crucially relying on the restrictions imposed by the outcome model: it returns a meaningful estimand that is directly informative about the conditional association between $A$ and $L$, even when model (\ref{linear}) is misspecified. Even so, caution is warranted as the restrictions of model (\ref{linear}) may be invoked when estimating $\omega(L)$ (e.g., based on smoothing splines under model (\ref{linear})), which may in turn may bias the assessment of the variability of $\hat{\beta}$. In particular, it may result in overly optimistic inferences about the conditional association between $A$ and $Y$, given $L$. This is indeed the case. Standard inference is based on standard errors estimated as the sample standard deviation of the so-called influence function of $\hat{\beta}$ under model (\ref{linear}):
\[\frac{\left\{A_i-\hat{\pi}(L_i)\right\}\left\{Y_i-\hat{\beta}A_i-\hat{\omega}(L_i)\right\}}{n^{-1/2}\sum_{i=1}^n \left\{A_i-\hat{\pi}(L_i)\right\}A_i}\]
\citep{robins1992estimating}. These ignore that when model (\ref{linear}) is misspecified, then different choices of $\pi(L)$ in (\ref{eq:estimand}) return estimands of a possibly different magnitude. This explains why excess variability may be observed when repeated samples deliver different estimates of $\pi(L)$. Moreover, as we will see in Section 5, under model misspecification $\hat{\pi}(L_i)$ will contribute to the first-order bias of the E-estimator, which is worrying when $\hat{\pi}(L_i)$ converges (in terms of root mean squared error) at a rate slower than $n^{-1/2}$. This is typically the case when smoothing splines are used, and is such that $\hat{\omega}(L_i)$ will then dominate the behaviour of the E-estimator.

In a simulation study under the above data-generating mechanisms, we found the empirical standard deviation of the E-estimator to be 16.7\% larger than estimated, resulting in 87.3\% coverage of 95\% confidence intervals for (\ref{eq:estimand}), despite the lack of bias in $\hat{\beta}$. In contrast, the nonparametric approach that we will develop later in this article, resulted in estimators with similar bias, and empirical standard deviation of the E-estimator being only 3.0\% larger than estimated (and being only 2.6\% larger than that of the E-estimator), resulting in 94.9\% coverage of 95\% confidence intervals for (\ref{eq:estimand}), despite the small sample size ($n=50$).

\section{Main effect estimands}

Suppose that interest lies in the association between a possibly continuous variable or exposure $A$ and a continuous outcome $Y$, conditional on measured variables $L$. One logical starting point would be the generalised partially linear model
\begin{equation}\label{gplm}
g\{E(Y|A,L)\}=\beta A+\omega(L),\end{equation}
where $g(\cdot)$ is a known link function and $\beta$ and $\omega(L)$ are unknown. This choice reflects the fact that in many regression analyses only a subset of the parameters are of scientific interest, and an analyst may prefer to be agnostic about the nuisance parameters. Model (\ref{gplm}) assumes a linear association as well as the absence of $A$-$L$ interactions (on the scale of the link function).
It does so for reasons of parsimony, e.g. because we may want to summarise the association between $A$ and $Y$ into a single number, 
but not because it reflects the ground truth. The general question, which we will work out in this paper, is then how to develop inference for $\beta$ in a way that does not rely on these assumptions.

The starting point of such analysis is to come up with an estimand that is meaningful when the above model does not hold, but reduces to $\beta$ when the model holds; this then subsequently allows for nonparametric inference to be developed for that estimand. One simple and generic strategy, which is sometimes advocated (e.g., \cite{van_der_laan_targeted_2011,buja_models_2019}), would be to define the estimand as a `projection' of the actual data distribution onto the (semi-parametric) model, such as the maximiser of the population expectation of the loglikelihood. This suggestion is useful, but vague as there will often be infinitely many such projection estimands. Indeed, each consistent estimator under the (semi)parametric model maps into a projection estimand, being defined as its probability limit under the nonparametric model. 

This calls for guidance concerning the choice of estimand in practice. In our development below, we will use several criteria for choosing an estimand.  Firstly, when the parametric assumptions hold, it should reduce to the target parameter of interest, in this case the parameter $\beta$ indexing (\ref{gplm}), to assure that the proposal does not hinder a familiar interpretation of the final result. Second, it should be generic, in the sense of being well defined regardless of whether $A$ is continuous or discrete. Indeed, the fact that parametric methods can flexibly incorporate any type of regressor no doubt contributes to their continuing appeal. It should also be generic in the sense that its influence curve should not demand the modelling of a (conditional) density, as flexible machine learning techniques are currently not well-adapted to density estimation, and this might make results very sensitive to the choice of density estimator. This criterion distinguishes our development from related work in the causal inference literature, where focus is usually given to binary exposures and effect modifiers.
 Third, the estimand must equal some $L$-dependent weighted average of the estimand one would choose to report for a subset of individuals with given $L$ (e.g. of the average outcome difference between subjects with $A=1$ versus $A=0$ and the same level of $L$). This ensures that the estimand captures what one is aiming for (e.g., a conditional association), which was not the case for ordinary least squares in Section 
\ref{sec:ill}. 




To distinguish assumptions aimed at parsimony from other, more substantive assumptions, let us 
start assuming that the main difficulty of the problem had already been solved.
Suppose in particular we already knew $E(Y|A=a,L)$ for all levels $a$ in the support of $A$ 
and all covariate levels $L$ over the support of $L$. 
Then we would generally not be interested in reporting exactly how $E(Y|A=a,L)$ changes over $a$ and $L$.
We would content ourselves with a parsimonious summary of the exposure effect.
At each level of $L$, a useful summary would be the conditional covariance between $A$ and $g\left\{E(Y|A,L)\right\}$, given $L$. 
This reduces to 
\[\pi(L)\left\{1-\pi(L)\right\}\left[g\left\{E(Y|A=1,L)\right\}-g\left\{E(Y|A=0,L)\right\}\right],\]
when $A$ is dichotomous (coded 0 or 1), where $\pi(L)$ is the so-called propensity score. This is clearly capturing a summary of the conditional association between $A$ and $Y$, given $L$, regardless of whether some model holds. This $L$-specific estimand can next be summarised across levels of $L$ as 
\begin{equation}\label{estimand-main}
\frac{E\left(\mbox{\rm Cov}\left[A,g\left\{E(Y|A,L)|L\right\}\right]\right)}{E\left\{\mbox{\rm Var}\left(A|L\right)\right\}}.\end{equation}
The denominator is here chosen to ensure that it reduces to $\beta$ under model (\ref{gplm}), but it remains unambiguously defined when this model is misspecified. For instance, it equals 
\[\frac{E\left(\pi(L)\left\{1-\pi(L)\right\}\left[g\left\{E(Y|A=1,L)\right\}-g\left\{E(Y|A=0,L)\right\}\right]\right)}{E\left[\pi(L)\left\{1-\pi(L)\right\}\right]},\]
when $A$ is dichotomous. It will therefore enable us to do inference for $\beta$ in model (\ref{gplm}) without relying on this model restriction. Interpretation of $\beta$ can still be done in the familiar way, relating to model (\ref{gplm}), however. But with the additional assurance that it continues to represent a summary of the conditional association between $A$ and $Y$, given $L$, when that model is misspecified; such assurance is not attained for standard maximum likelihood estimators, for instance, as we saw in Section \ref{sec:ill}.

The estimand (\ref{estimand-main}) with $g(.)$ the identity link has been studied by a number of authors, e.g. \cite{robins2008higher,newey_cross-fitting_2018,whitney2019}. We will here extend inference for it to arbitrary link functions. Such extension is non-trivial, if one considers the major difficulties that have been experienced in drawing inference for $\beta$ under the partially linear logistic model \citep{tchetgen2010doubly,tan2019doubly}, which have resulted in elegant, but complex proposals that require the modelling of the conditional density or mean of the exposure, given outcome and covariates; relying on such models is arguably less desirable when information about the conditional density of the exposure, given covariates but not outcome, is a priori available (as in randomised experiments, for instance). 
These complications will be avoided with our choice of estimand (\ref{estimand-main}), which also reduces to $\beta$ under model (\ref{gplm}) with $g(.)$ the logit link, for which we develop nonparametric inference in Section \ref{sec:inference}. 
This extension is moreover important since the probability limits of popular estimators of parameters indexing non-linear models have no simple closed-form representation (unlike was the case for the OLS estimator in Section \ref{sec:ill}), thus rendering their behaviour ill understood when the model restrictions fail to hold. In particular, estimators for $\beta$ based on the semiparametric efficient score will generally fail to converge to (\ref{estimand-main}).


When the exposure is dichotomous (taking values 0 and 1),  $g(.)$ is the identity link and moreover $L$ is sufficient to adjust for confounding (in the sense that $A$ is independent of the counterfactual outcome $Y^a$ to exposure level $a$, given $L$), then (\ref{estimand-main}) reduces to 
\begin{equation}\label{overlap-linear}
\frac{E\left[\pi(L)\left\{1-\pi(L)\right\}(Y^1-Y^0)\right]}{E\left[\pi(L)\left\{1-\pi(L)\right\}\right]}.\end{equation}
This effect, which was also considered in \cite{crump2006moving} and \cite{vansteelandt2014regression}, gives highest weighted to covariate regions where both treated and untreated subjects are found. It expresses the exposure effect that would be observed in a randomised experiment where the chance of recruitment is proportional to both the probability of being treated as well as the probability of being untreated. In that case, subjects with a 10\% chance of receiving treatment (or no treatment) are roughly 10 times more likely to be recruited than subjects with a 1\% chance of receiving treatment (or no treatment), while subjects whose chance of receiving treatment lies between 25\% and 75\% are nearly equally likely to be recruited (their chance of recruitment deviates at most 33\% in relative terms). Although such recruitment probabilities are not readily applied in a real-life setting, the resulting effect may well approximate that which would be found in a real-life randomised experiment, where the eligibility criteria would exclude patients who are unlikely to receive treatment or no treatment in practice. Regarding the optimality properties of this estimand, \cite{crump2006moving} consider the class of weighted sample average treatment effects $\sum^n_{i=1}w(L_i)(Y^1_i-Y^0_i)/\sum^n_{i=1}w(L_i)$ where $w(L)$ is a (known) weight. They show that the choice $w(L)=\pi(L)\{1-\pi(L)\}$ delivers the parameter that can be estimated with the greatest precision across the entire class. 

The estimand (\ref{estimand-main}) thus generalises the propensity-overlap-weighted effects to more general exposures and arbitrary link functions.
Such generalisation becomes essential when the exposure is continuous, in view of the need to summarise the (now high-dimensional) exposure effect.

\section{Effect modification estimands}

Suppose next that interest lies in the interaction between two possibly continuous variables $A_1$ and $A_2$ on a continuous outcome $Y$, conditional on measured variables $L$. One logical starting point is the partially linear interaction model \citep{vansteelandt2008multiply}
\begin{equation}\label{glm_int}
g\{E(Y|A_1,A_2,L)\}=\omega_1(A_1,L)+\omega_2(A_2,L)+\beta A_1A_2,\end{equation}
where $\beta,\omega_1(A_1,L)$ and $\omega_2(A_2,L)$ are unknown. The construction of a generic estimand that reduces to $\beta$ when model (\ref{glm_int}) is correctly specified, turns out a non-trivial task. We are not aware of existing estimands for interaction parameters that satisfy the criteria in Section 3; even if were to accept parameters whose influence curve requires modelling a density, current proposals are limited to binary $A_1$ and $A_2$ \citep{van_der_laan_targeted_2011}.

Let us therefore first consider the case where $A_1$ and $A_2$ are dichotomous.
At each level of $L$, a useful interaction summary would be
\[\mu_{11}(L)+\mu_{00}(L)-\mu_{10}(L)-\mu_{01}(L),\]
where $\mu_{a_1a_2}(L)\equiv g\{E(Y|A_1=a_1,A_2=a_2,L)\}$.
Summarising across levels of $L$, we may then consider the estimand 
\begin{equation}\label{estimandint-linear}
\frac{E\left[\pi_1(L)\left\{1-\pi_1(L)\right\}\pi_2(L)\left\{1-\pi_2(L)\right\}\left\{\mu_{11}(L)+\mu_{00}(L)-\mu_{10}(L)-\mu_{01}(L)\right\}\right]}{E\left[\pi_1(L)\left\{1-\pi_1(L)\right\}\pi_2(L)\left\{1-\pi_2(L)\right\}\right]},\end{equation}
where $\pi_1(L)=P(A_1=1|L)$ and $\pi_2(L)=P(A_2=1|L)$.
It assigns highest weight to subjects for whom each exposure combination is sufficiently likely, so as to avoid extrapolation towards covariate strata that carry little or no information about interaction. This estimand reduces to $\beta$ under model (\ref{glm_int}), but by construction, continues to represent a weighted average of $L$-specific interactions when that model is misspecified; remember that we have no such guarantee with standard estimation approaches for interactions. 
When $L$ is sufficient to adjust for confounding for the effect of both exposures (in the sense that $(A_1,A_2)$ is independent of the counterfactual outcome $Y^{a_1a_2}$ to exposure $(a_1,a_2)$, given $L$) and $g(\cdot)$ is the identity link, then estimand (\ref{estimandint-linear}) can also be written as
\begin{equation}\label{overlap-intlinear}
\frac{E\left[\pi_1(L)\left\{1-\pi_1(L)\right\}\pi_2(L)\left\{1-\pi_2(L)\right\}(Y^{11}-Y^{10}-Y^{01}+Y^{00})\right]}{E\left[\pi_1(L)\left\{1-\pi_1(L)\right\}\pi_2(L)\left\{1-\pi_2(L)\right\}\right]}.\end{equation}

Extending (\ref{estimandint-linear}) to arbitrary exposures is challenging. A natural generalisation is
\begin{equation}\label{estimandint-linearg}
\frac{E\left[\left\{A_1-E(A_1|L)\right\}\left\{A_2-E(A_2|L)\right\}g\{E(Y|A_1,A_2,L)\}\right]}{E\left[\left\{A_1-E(A_1|L)\right\}^2\left\{A_2-E(A_2|L)\right\}^2\right]},\end{equation}
which reduces to $\beta$ when model (\ref{glm_int}) is correctly specified, as well as to (\ref{estimandint-linear}) when both exposures are dichotomous and conditionally independent, given $L$.
Extending (\ref{estimandint-linearg}) further to the case where $A_1$ and $A_2$ may be conditionally dependent, given $L$, brings additional challenges.
While in principle the estimand (\ref{estimandint-linearg}) still meaningfully summarises the degree of effect modification in the case of conditionally dependent exposures, distribution-free inference for it demands inverse weighting by the joint density of both exposures, conditional on $L$, and this makes generic inference for it rather challenging. The reason for this is best understood in the case of dichotomous exposures. Here, (\ref{estimandint-linear}) does not downweigh covariate strata where for instance subjects with $A_1=1$ and $A_2=1$ are extremely rare, despite each level of $A_1$ and each level of $A_2$ being well represented. In such cases, the estimand (\ref{estimandint-linear}) necessitates extrapolations away from the observed exposure distribution, and this in turn complicates inference.  
For dichotomous exposures, a natural generalisation of (\ref{estimandint-linear}) which overcomes the previous concerns is
\[\frac{E\left[\frac{\pi_{11}(L)\pi_{10}(L)\pi_{01}(L)\pi_{00}(L)}{\pi_1(L)\left\{1-\pi_1(L)\right\}\pi_2(L)\left\{1-\pi_2(L)\right\}}\left\{\mu_{11}(L)+\mu_{00}(L)-\mu_{10}(L)-\mu_{01}(L)\right\}\right]}{E\left[\frac{\pi_{11}(L)\pi_{10}(L)\pi_{01}(L)\pi_{00}(L)}{\pi_1(L)\left\{1-\pi_1(L)\right\}\pi_2(L)\left\{1-\pi_2(L)\right\}}\right]},\]
where $\pi_{a_1a_2}(L)\equiv P(A_1=a_1,A_2=a_2|L)$ for $a_1,a_2=0,1$. This estimand reduces to (\ref{overlap-intlinear}) under conditional independence, but it is unclear how it can be generalised to arbitrary exposures. In view of this, we instead choose to work with the following estimand:
\begin{equation}\label{estimandintg-linear}
\frac{E\left[P(A_1A_2)g\left\{E(Y|A_1,A_2,L)\right\}\right]}{E\left[P(A_1A_2)^2\right]},\end{equation}
where $P(.)$ is an orthogonal projection operator (w.r.t. the covariance inner product), which projects an arbitrary function of $(A_1,A_2,L)$ onto the space of the functions of $(A_1,A_2,L)$ with mean zero, conditional on $A_1,L$ as well as conditional on $A_2,L$. Such projection eliminates from $g\left\{E(Y|A_1,A_2,L)\right\}$ all main effects of $A_1$ and $L$ (as well as their interactions) and all main effects of $A_2$ and $L$ (as well as their interactions), thus leaving only its dependence on functions of both $A_1$ and $A_2$ (and $L$) that cannot be additively separated into functions of $(A_1,L)$ or $(A_2,L)$; such functions define additive interactions between $A_1$ and $A_2$ on the scale of the link function $g(.)$. This is best understood for dichotomous exposures, where (\ref{estimandintg-linear}) reduces to a weighted average of $L$-conditional interactions. Indeed, for such exposures we can always write
\begin{eqnarray*}
g\left\{E(Y|A_1,A_2,L)\right\}&=&c_0(L)+c_1(L)A_1+c_2(L)A_2\\
&&+\left\{\mu_{11}(L)+\mu_{00}(L)-\mu_{10}(L)-\mu_{01}(L)\right\}A_1A_2,\end{eqnarray*}
for certain functions $c_j(L),j=1,2,3$.
This and the fact that $c_0(L)+c_1(L)A_1+c_2(L)A_2$ is orthogonal (w.r.t. the covariance inner product) to $P(A_1A_2)$ implies that the estimand reduces to 
\[\frac{E\left[P(A_1A_2)^2\left\{\mu_{11}(L)+\mu_{00}(L)-\mu_{10}(L)-\mu_{01}(L)\right\}\right]}{E\left\{P(A_1A_2)^2\right\}}.\]
More generally, (\ref{estimandintg-linear}) reduces to (\ref{estimandint-linear}) when $A_1$ and $A_2$ are conditionally independent, given $L$.

\section{Nonparametric inference}\label{sec:inference}

In the previous sections, we have shown how modelling assumptions can be invoked to summarise the (conditional) association between two variables, which may itself be high-dimensional, or the extent to which two variables interact in their effect on outcome. To prevent that these convenience assumptions are used as a ground truth, we next develop inference for the resulting estimands under a nonparametric model. 

\subsection{The influence curve}

Inference under a nonparametric model is based on the use of so-called influence curves \citep{pfanzagl1990estimation,bickel1993efficient}.
Technically, this is mean zero functional of the observed data and the data-generating distribution, which 
characterises the estimand's sensitivity to arbitrary (smooth) changes in the data-generating law.
The estimand (\ref{estimand-main}), for instance, has an influence curve given by the following expression
\begin{align}\label{robinson}
\frac{\left\{A-E(A|L)\right\}\left[\mu(Y,A,L)-\beta \left\{A - E(A|L)\right\}\right]}{E\left[\left\{A-E(A|L)\right\}^2\right]}
\end{align}
(see the appendix), where $\beta$ is given by (\ref{estimand-main}) and 
\[\mu(Y,A,L)=g'\{E(Y|A,L)\}\{Y-E(Y|A,L)\}+g\{E(Y|A,L)\}-E[g\{E(Y|A,L)\}|L]\] and $g'(x)=\partial g(x)/\partial x$.
If the conditional expectations indexing the influence curve were known, then it would follow from its mean zero property that a consistent estimator $\hat{\beta}$ of $\beta$ could be obtained as the value of $\beta$ that makes the sample average of the influence curves zero. The resulting estimator's asymptotic distribution would be governed by this influence curve in the sense that
\[\sqrt{n}\left(\hat{\beta}-\beta\right)=\frac{1}{\sqrt{n}}\sum_{i=1}^n \frac{\left\{A_i-E(A|L_i)\right\}\left[\mu(Y_i,A_i,L_i)-\beta \left\{A_i - E(A|L_i)\right\}\right]}{E\left[\left\{A-E(A|L)\right\}^2\right]}+o_p(1).\]
The fact that the difference between the estimator and the truth can be approximated by a sample average of functions, known as influence functions, implies that $\hat{\beta}$ is asymptotically linear with influence function given by the influence curve. This implies in turn that it is asymptotically normally distribution with bias that shrinks to zero faster than the standard error, and with a variance that can be estimated as the sample variance of the influence functions (where population expectations and the value of $\beta$ can be substituted by consistent estimates). In the special case where $g(.)$ is the identity link, the resulting estimator thus has influence function given by 
\[\frac{\left\{A-E(A|L)\right\}\left[Y-E(Y|L)-\beta \left\{A - E(A|L)\right\}\right]}{E\left[\left\{A-E(A|L)\right\}^2\right]},\]
which interestingly is the influence function for the `partialling-out' estimator proposed by \cite{robinson1988root} for the parameter indexing the partially linear model (\ref{linear}). Hence by following the proposal above, we obtain a generalised `partialling-out' estimator for $\beta$. 


The fact that the influence curve involves unknown conditional expectations, makes the estimator $\hat{\beta}$ suggested in the previous paragraph infeasible. 
In practice, we will therefore substitute these by consistent estimators. 
Interestingly, provided that these converge sufficiently fast in a relatively weak sense made specific in the Appendix, the resulting estimator of $\beta$ behaves the same as if these conditional expectations were given and known. In particular, its variance can be estimated as previously suggested, namely as 1 over $n$ times the sample variance of the influence curves, as if these conditional expectations were given. It makes the use of influence curves extremely powerful, as it implies in particular that the uncertainty that the estimators of the conditional expectations add to the analysis can be ignored when drawing inference about $\beta$, even when these are based on variable selection or machine learning procedures, whose uncertainty is difficult to quantify; see the next section.


It is instructive to contrast inference, as described above, for the estimand (\ref{estimand-main}), versus inference for the E-estimator \citep{robins1992estimating} of $\beta$ under model (\ref{linear}). The latter is a specific G-estimator, which 
converges to (\ref{estimand-main}), even when model (\ref{linear}) is misspecified \citep{vansteelandt2014regression}.
Its influence function
\[\frac{\left\{A-E(A|L)\right\}\left\{Y-\beta A - E(Y|A=0,L)\right\}}{E\left[\left\{A-E(A|L)\right\}^2\right]},\]
coincides with the above influence curve when model (\ref{linear}) holds, but not necessarily otherwise.
The implications of this are best appreciated when specialising to a dichotomous exposure and considering the corresponding estimand (\ref{overlap-linear}).
When model (\ref{linear}) is misspecified, then each change of $\pi(L)$ also changes the estimand. In particular, different estimates of the propensity score may then be viewed as targeting different effect estimands. The resulting excess variability is not acknowledged when basing inference on the influence function of the E-estimator, as this is assuming model (\ref{linear}) to be correctly specified. This was indeed what was observed in the simulation study described in Section 2. As \cite{buja2019} note, for certain choices of nuisance parameter estimators (specifically, series methods or twicing kernels) the E-estimator and the proposed influence-curve estimator can exactly coincide. However, since we wish to work in greater generality, and in the following section consider arbitrary machine learners for  the nuisances, we do not consider this subtlety any further.

\subsection{Implementation using machine learning}

\subsubsection{Main effect estimands}

In order to construct an estimator of $\beta$, given by (\ref{estimand-main}), we must first obtain estimates of the quantities $E(A|L)$, $E(Y|A,L)$ and $E[g\{E(Y|A,L)\}|L]$. Since we wish to be `assumption-free' (or at least, assumption-lean) it is natural to want to do this without pre-specification of parametric models. One could therefore adopt variable/model selection procedures, or use traditional nonparametric estimators (e.g. kernel methods, sieve estimators, regression trees) or even machine learning approaches (random forests, neural networks, support vector machines) which are particularly effective when the dimension of the covariates is large. Machine learning techniques learn a (potentially very complex) `model' from the data, whilst using regularisation (in combination with cross-validation) to minimise issues of overfitting and optimise out-of-sample predictive performance. The analyst does not need choose between different estimators now available in statistical software; ensemble learners (such as the Super Learner) aim to take the optimal weighted combination of candidate (parametric and nonparametric) estimators.

Traditionally, statisticians have been hesitant to routinely incorporate machine learning when analysing data. To illustrate why, if we return to the running example of of inference in model (\ref{linear}), suppose that in a first step we obtain an estimate of $\hat{E}(A|L)$ 
using data adaptive methods, which are used to construct an estimator of $\beta$ as
\begin{align}\label{EE}
0=\frac{\sum^n_{i=1}\left\{A_i-\hat{E}(A_i|L_i)\right\}Y_i}{\sum^n_{i=1}\left\{A_i-\hat{E}(A_i|L_i)\right\}A_i}.
\end{align}
Such an estimator appears very natural, given that we are essentially solving the sample analogue of the population equations 
\[0=E\left[\left\{A-{E}(A|L)\right\}\left\{E(Y|A,L)-\beta A\right\}\right]=E\left[\left\{A-{E}(A|L)\right\}(Y-\beta A)\right]\]
which have a solution at the estimand $\beta$, except that estimates of the unknown conditional expectation are `plugged in'. The tuning parameters used to control the degree of regularisation in the first step estimator  are typically chosen to balance bias and variance in a way that is optimal for prediction purposes. Unfortunately, this choice is usually \textit{suboptimal} for estimation of the target parameter; the bias of na\"ive `plug-in' estimator of $\beta$ depends on the error $\hat{E}(A|L)-{E}(A|L)$ and thus can inherit the potentially large bias in $\hat{E}(A|L)$. The consequence is that the bias of the naive estimator may be of the order $n^{1/2}$ or larger, and hence the use of standard confidence intervals is not justified.

So long as the partially linear model (\ref{linear}) holds, it turns out that there are several different ways of constructing estimators of $\beta$ that are desensitised to `plug-in' bias of machine learners. \citet{chernozhukov_double/debiased_2018} propose using either the E-estimator (\ref{eestimator}) or the `partialling out' estimator of \citet{robinson1988root}
\begin{align}\label{robinson_88}
\frac{\sum^n_{i=1}\left\{A_i-\hat{E}(A_i|L_i)\right\}\{Y_i-\hat{E}(Y_i|L_i)\}}{\sum^n_{i=1}\left\{A_i-\hat{E}(A_i|L_i)\right\}^2}.
\end{align}
where all nuisance parameter are estimated via machine learning. The asymptotic bias of both of these approaches depends in part on the product of two errors  -  either
\[\{E(A|L)-\hat{E}(A|L)\}\{E(Y|A=0,L)-\hat{E}(Y|A=0,L)\}\] for the E-estimator or 
\begin{align}\label{p_bias_robinson}
\{E(A|L)-\hat{E}(A|L)\}[E(Y|L)-\hat{E}(Y|L)-\beta\{E(A|L)-\hat{E}(A|L)\}]
\end{align} for the `partialling out' estimator. 
As long as each estimator converges to the truth, then the product of two errors will tend to shrink at least as fast (and usually much faster) than an individual error. Indeed, if each nuisance estimator converges at a rate faster than $n^{1/4}$, then the bias of the desensitised estimator is faster than $n^{-1/2}$, enabling parametric-rate inference (see the Appendix for further details).

So long as the semiparametric model restriction holds, both estimation approaches discussed in the previous paragraph are first-order equivalent. However, the situation is quite different when the restriction fails \citep{whitney2019}. For the estimand (\ref{estimand-main}), the asymptotic bias of the estimator is now proportional to
\[E\left[\left\{E(A|L)-\tilde{E}(A|L)\right\}\left\{E(Y-\beta A|L)-\tilde{E}(Y|A=0,L)\right\}\right],\]
where $\tilde{E}(A|L)$ is the probability limit of $\hat{E}(A|L)$ and $\tilde{E}(Y|A=0,L)$ is the probability limit of $\hat{E}(Y|A=0,L)$;
note that $E(Y|A=0,L)=E(Y-\beta A|L)$ under the partially linear model but not otherwise. Because the error $E(Y-\beta A|L)-\hat{E}(Y|A=0,L)$ will no longer shrink to zero, the bias of the E-estimator will be determined by $E(A|L)-\hat{E}(A|L)$. As discussed above, the situation may be much worse for semiparametric estimators in nonlinear models, since the bias w.r.t (\ref{estimand-main}) may now even diverge. By considering (\ref{p_bias_robinson}), it follows that the same issues are not true for the `partialling out' estimator, which makes the sample average of the influence curves for the estimand (\ref{estimand-main}) evaluated at the 
machine learning predictions equal to zero. This highlights the benefits of estimation using the influence curve obtained under a nonparametric model; it incorporates an implicit bias-correction, as the bias of the estimator of the target parameter is usually smaller in magnitude than that of the first stage estimators. Moreover, this property is not dependent on any semiparametric modelling assumptions.

Hence after deriving the influence curve, we can obtain a estimator and confidence interval by following the simple recipe below: 
\begin{enumerate}
\item  Obtain the estimates $\hat{E}(A|L)$ and $\hat{E}(Y|A,L)$, e.g. using machine learning.
\item If $A$ is binary, estimate $E[g\{E(Y|A,L)\}|L]$ as
\[
\hat{E}[g\{\hat{E}(Y|A,L)\}|L]=g\{\hat{E}(Y|A=1,L)\} \hat{E}(A|L)+g\{\hat{E}(Y|A=0,L)\}\{1-\hat{E}(A|L)\}
\]
otherwise, use an additional machine learning fit (with $g\{\hat{E}(Y|A,L)\}$ as outcome). 
\item Obtain an estimate of $\mu(Y,A,L)$:
\begin{eqnarray*}
\hat{\mu}(Y,A,L)&=&g^{-1}\{\hat{E}(Y|A,L)\}\{Y-\hat{E}(Y|A,L)\}\\
&&+g\{\hat{E}(Y|A,L)\}-\hat{E}[g\{\hat{E}(Y|A,L)\}|L].
\end{eqnarray*}
\item  Fit a linear regression of $\hat{\mu}(Y,A,L)$ on the sole predictor $A-\hat{E}(A|L)$ (without an intercept) using OLS in order to obtain an estimate $\hat{\beta}$ of $\beta$. 
\end{enumerate}
Remarkably, a valid standard error can be obtained by requesting that the regression software uses a sandwich estimator. This is because the influence curves are already `pre-orthogonalised' or `de-sensitised' with respect to nuisance parameter estimators. When used in combination with cross-fitting (see the next paragraph), one can therefore just look at the empirical variance of the influence curve, in the same way one would calculate the variance of a sample mean. 

In order for the resulting confidence interval to be valid, some additional assumptions are required. In addition to the aforementioned rate conditions on $\hat{E}(A|L)$, $\hat{E}(Y|A,L)$ and $\hat{E}[g\{\hat{E}(Y|A,L)\}|L]$ (which are made more specific in the Appendix), we also need to restrict the complexity of these estimators. Historically, this complexity has often been bounded using empirical process conditions, but these are unlikely to be satisfied for very flexible machine learning methods. A simple solution is to use sample-splitting; split the data in half, estimate the nuisance parameters in the `training' split, and perform inference one $\beta$ in the `validation' sample. This has a disadvantage of halving the sample size. However, efficiency can be asymptotically recovered via cross-fitting \citep{van_der_laan_cross-validated_2011,chernozhukov_double/debiased_2018}; e.g. one can reverse the training and validation samples, construct a second estimate of $\beta$ and average the pair. The variance can estimated by combining the predicted influence curves (used to estimate $\beta$) from each split, replacing $\beta$ with the averaged rather than split-specific estimate; one can then take the sample average of the curves as before. 


The combination of influence curve-based estimators with cross-fitting facilitates the use of machine learning to estimate parts of the data distribution of no scientific interest. These important results have only been highlighted relatively recently, and many open questions remain. Firstly, there is yet to be firm guidance on the number of splits to use in the cross-fitting. Moreover, since the machine learning methods typically perform better with more data, it may be that no splitting can sometimes yield superior performance to cross-fitting. At the other extreme, due to the similarity of our estimator to that of \cite{robinson1988root}, it may be possible to obtain much sharper results on the nuisance estimators by using a more specific variant of cross-fitting in combination with so-called `undersmoothing' \citep{newey_cross-fitting_2018}. This is left to future work. For now, if cross-fitting is adopted, we recommend 10-fold cross-fitting, each time using nine tenths as training sample and the remainder as validation sample.

\subsubsection{Effect modification estimands}

Let us first consider inference for the effect modification estimand $\beta$, given by (\ref{estimandint-linear}), 
under the assumption of conditionally independent exposures.
In the Appendix, we show that its influence curve under the nonparametric model equals
\begin{eqnarray*}
&&-\left(E\left[\left\{A_1-E(A_1|L)\right\}\left\{A_2-E(A_2|L)\right\}A_1A_2\right]\right)^{-1}\\
&&\times \left(\left\{A_1-E(A_1|L)\right\}\left\{A_2-E(A_2|L)\right\}g'\{E(Y|A_1,A_2,L)\}\left\{Y-E(Y|A_1,A_2,L)\right\}\right.\\
&&\left. + \left\{A_1-E(A_1|L)\right\}\left\{A_2-E(A_2|L)\right\} \left[g\{E(Y|A_1,A_2,L)\}-\beta A_1A_2\right]\right. \\
&&\left. - E\left(\left\{A_2-E(A_2|L)\right\}\left[g\{E(Y|A_1,A_2)\}-\beta A_1A_2\right]|L\right) \left\{A_1-E(A_1|L)\right\}\right.\\
&&\left. - E\left(\left\{A_1-E(A_1|L)\right\}\left[g\{E(Y|A_1,A_2)\}-\beta A_1A_2\right]|L\right) \left\{A_2-E(A_2|L)\right\}\right).
\end{eqnarray*}
A root-$n$ consistent estimator of $\beta$ can thus be obtained by solving an estimating equation with estimating function given by the influence curve. That is,
\begin{eqnarray*}
\hat{\beta}&=&\frac{1}{n}\sum_{i=1}^n \left(\left\{A_{i1}-\hat{E}(A_1|L_i)\right\}\left\{A_{i2}-\hat{E}(A_2|L_i)\right\}g'\{\hat{E}(Y|A_{i1},A_{i2},L_i)\}\left\{Y-\hat{E}(Y|A_{i1},A_{i2},L_i)\right\}\right.\\
&& \left.+\left\{A_{i1}-\hat{E}(A_1|L_i)\right\}\left\{A_2-\hat{E}(A_2|L_i)\right\}  g\{\hat{E}(Y|A_{i1},A_{i2},L_i)\}  \right.\\
&& \left.- \hat{E}\left[\left\{A_2-\hat{E}(A_2|L_i)\right\}g\{\hat{E}(Y|A_{i1},A_{i2},L_i)\}|L_i\right] \left\{A_{i1}-\hat{E}(A_1|L_i)\right\}\right.\\
&& \left.- \hat{E}\left[\left\{A_1-\hat{E}(A_1|L_i)\right\}g\{\hat{E}(Y|A_{i1},A_{i2},L_i)\}|L_i\right] \left\{A_{i2}-\hat{E}(A_2|L_i)\right\}\right)\\
&&\times \left(\frac{1}{n}\sum_{i=1}^n \left[\left\{A_{i1}-\hat{E}(A_1|L_i)\right\}\left\{A_{i2}-\hat{E}(A_2|L_i)\right\}A_{i1}A_{i2}\right.\right.\\
&& \left.\left.- \hat{E}\left[\left\{A_2-\hat{E}(A_2|L_i)\right\}^2|L_i\right] \hat{E}(A_1|L_i)- \hat{E}\left[\left\{A_1-\hat{E}(A_1|L_i)\right\}^2|L_i\right] \hat{E}(A_2|L_i)\right]\right)^{-1},
\end{eqnarray*}
where $\hat{E}(A_j|L_i),\hat{E}\left[\left\{A_j-\hat{E}(A_j|L_i)\right\}g\{\hat{E}(Y|A_{1},A_{2},L_i)\}|L_i\right],\hat{E}\left[\left\{A_j-\hat{E}(A_j|L_i)\right\}^2|L_i\right]$ for $j=1,2$ and $g\{\hat{E}(Y|A_{1},A_{2},L_i)\}$ denote machine learning predictions. 

Consider next the effect modification estimand $\beta$, given by (\ref{estimandintg-linear}), without the assumption of conditional independence.
In the Appendix, we show that its influence curve under the nonparametric model equals
\[\frac{P(A_1A_2)}{E\left\{P(A_1A_2)^2\right\}}\left(g'\{E(Y|A_1,A_2,L)\}\{Y-E(Y|A_1,A_2,L)\}+P\left[g\{E(Y|A_1,A_2,L)\}-\beta A_1A_2\right]\right)\]
A root-$n$ consistent estimator of $\beta$ can thus be obtained as
\begin{eqnarray*}
\hat{\beta}&=&\frac{1}{n}\sum_{i=1}^n \hat{P}(A_{i1}A_{i2})\left(g'\{\hat{E}(Y|A_{i1},A_{i2},L_i)\}\{Y_i-\hat{E}(Y|A_{i1},A_{i2},L_i)\}\right.\\&&\left.+\hat{P}\left[g\{\hat{E}(Y|A_{i1},A_{i2},L_i)\}\right]\right)\left\{\frac{1}{n}\sum_{i=1}^n \hat{P}^2(A_{i1}A_{i2})\right\}^{-1}.
\end{eqnarray*}
Here, $\hat{E}(Y|A_1,A_2,L_i)$ denotes a machine learning prediction. Further, the projection $\hat{P}(A_{i1}A_{i2})$
can be obtained via the alternating conditional expectations (ACE) algorithm \citep{bickel1993efficient}. This involves first predicting $A_{i1}A_{i2}$ on the basis of $A_{i1}$ and $L_i$ via machine learning and taking the residuals; next, predict these residuals on the basis of $A_{i2}$ and $L_i$ via machine learning and take the residuals; next, predict these residuals on the basis of $A_{i1}$ and $L_i$ via machine learning and take the residuals; and so forth. This process can be aborted when the variance of the predicted residuals reaches a value very close to zero. To ensure a decreasing variance, we recommend in each step tuning the obtained machine learning predictions of the residuals by substituting these by the ordinary least squares prediction of those residuals onto the obtained machine learning predictions. The projection $\hat{P}\left\{\hat{E}\left(Y|A_{i1},A_{i2},L_i\right)\right\}$ is likewise obtained, starting from $\hat{E}\left(Y|A_{i1},A_{i2},L_i\right)$.

The variance of both considered estimators is obtained as 1 over $n$ times the variance of the corresponding influence curve, with conditional expectations substituted by machine learning predictions, marginal expectations by sample averages and $\beta$ by $\hat{\beta}$.

\section{Simulation studies}

\subsection{Main effect estimands}

We considered inference based on the partially linear logistic model $\logit\{E(Y|A,L)\}=\beta A+\omega(L)$. Specifically, we generated a 10-dimensional covariate $L\sim N(0,\Sigma)$, where $\Sigma$ was 
randomly generated with variances between 2 and 10 and correlations up to 0.6 in absolute value; 
and $A\sim \Bern(\gamma^T L - 0.15L^2_1)$, where $\gamma$ is the 10-dimensional unit vector scaled by $1/40$ and $L_k$ is the $k$th entry of $L$. To investigate performance when the restriction does not hold, we considered 4 separate settings: 1) $Y\sim \Bern(\expit(0.3A+\delta^T L_{[1:5]}))$ where $\delta$ is a 5 unit vector scaled by $1/50$; 2) $Y\sim \Bern(\expit(0.3A+\delta^T L_{[1:5]}+0.1L^2_1))$; 3) $Y\sim \Bern(\expit(1.5L_1(A-1)+\delta^T L_{[1:5]}))$; and 4) $Y\sim \Bern(\expit(0+0.1/(1+\exp(0.25 L_3-0.25L_2))+0.3A/(1+\exp(-0.25L_2))+0.5AL_6+0.1L^2_1))$. Only in the first two settings does the model restriction hold; the fourth setting is especially challenging, in light of the complex functional form of the interaction between $A$ and $L$.

In the simulations, we included the maximum likelihood estimator (MLE) of $\beta$ obtained by fitting the logistic regression model $\logit\{E(Y|A,L)\}=\beta A+\alpha_0+\alpha^T_1 L$; this model was misspecified in experiments 2-4 due to the omitted quadratic effect of $L$. We also included two estimators designed for the partially linear logistic model; the first estimator `ES' solves the semiparametric efficient score equations e.g. in \citet{kosorok2007introduction}:
\begin{align*}
0=&\sum^n_{i=1}\left(A_i-\frac{\hat{E}\left[A_i\hat{E}(Y_i|A_i,L_i)\{1-\hat{E}(Y_i|A_i,L_i)\}\big|L_i\right]}{\hat{E}\left[\hat{E}(Y_i|A_i,L_i)\{1-\hat{E}(Y_i|A_i,L_i)\}\big|L_i\right]}\right)\\&\times\left(Y_i-\expit\left[\beta A_i+\logit\{\hat{E}(Y_i|A_i=0,L_i)\} \right]\right)
\end{align*}
whereas the second is the simple doubly robust (DR) estimator proposed in \citet{tchetgen2013closed}, which solves the equations
\begin{align*}
0=&\sum^n_{i=1}\left\{A_i-\hat{E}(A_i|Y_i=0,L_i)\right\}\left\{Y_i-\hat{E}(Y|A_i=0,L_i)\right\}\exp(-\beta A_i Y_i)
\end{align*}
Finally, we considered the proposal described in Sections 4 and 5. 
For the ES estimator as well as the proposal, random forests (via the `grf' package described in \cite{athey2019generalized}) were used to learn $E(Y|A,L)$ and $E(A|L)$ and yield predictions (as well as predictions of $E(Y|A=a,L)$ for $a=0,1$). These could then be plugged into the relevant estimating equations via application of the law of total probability. For the DR estimator, random forests were used to learn $E(A|Y,L)$ so that predictions of $E(A|Y=0,L)$ could be obtained, as this reflects how this conditional expectation would likely be estimated in practice using machine learning. In experiments 2-4 to conceptualise bias and coverage for the MLE, we generated 500 datasets with sample size 100,000 and took the average of the estimates obtained from a misspecified model as the limiting quantity targeted by the MLE. We did the same for the semi- and nonparametric methods except used the true conditional expectations for the nuisance functionals (since we assume the random forests converge to the truth). Comparing each estimator with its own limit meant that our proposed estimand was not privileged, and gives a fair reflection of the properties of each approach (e.g. a confidence interval may have poor coverage even for its own estimand). 


In Table \ref{main_difficult}, we see that the MLE does fairly well in terms of coverage, although the model-based estimator of the variance is usually an underestimate (this could potentially be remedied by using a sandwich estimator). However, there is no guarantee that the MLE targets an estimand that summarises the conditional association of scientific interest. This is confirmed in experiment 2, where the limit of the MLE is in a different direction to the parameter in the partially linear model, which is especially worrisome. 

We see that the two semiparametric approaches perform well when the model restriction holds; in experiments 3 and 4 however we see that outside of the model, coverage can sharply decrease as sample size increases. This is particularly the case for the DR estimator, where the plug-in bias inherited from the random forests appears to be substantial. Compared with these approaches, our proposal has the better coverage of the approaches considered across the different sample sizes; this is due to both lower bias, and estimated standard errors that at least in large samples more accurately reflect the variability of the estimator. Reassuringly, despite our inferences being assumption-lean, the empirical standard deviations show that this does not come with a loss of precision.

\begin{table}
\caption{\label{main_difficult} Simulation results on main effects: empirical bias (Bias) and standard deviation (Emp SD), sample average of the estimated influence-curve based standard errors (Mean SE), and coverage of 95\% Wald confidence intervals (Cov). Bias and coverage taken w.r.t. limiting values of each estimator:  0.3 (all estimators) in experiment 1; -0.28 (MLE) and 0.3 (ES, DR and proposal) in experiment 2; 0.33 (MLE) 0.43 (ES), 1.00 (DR) and 0.50 (proposal) in experiment 3 and 0.025 (MLE), 0.24 (ES) 0.38 (DR) and 0.23 (proposal) in experiment 4.}
\resizebox{\columnwidth}{!}
{
\begin{tabular}{lllllllllllllllllll}\hline
Exper. & $n$ & \multicolumn{4}{c}{MLE} & \multicolumn{4}{c}{ES} &  \multicolumn{4}{c}{DR}  & \multicolumn{4}{c}{Proposal} \\
& & Bias & ESD & MSE & Cov & Bias & ESD & MSE & Cov& Bias & ESD & MSE & Cov & Bias & ESD & MSE & Cov \\\hline
 1 & 500 & 0.00 & 0.21 & 0.21 & 95 & 0.04 & 0.20 & 0.19 & 93 & 0.06 & 0.21 & 0.23 & 96 & 0.02 & 0.19 & 0.20 & 95 \\ 
& 1000 & 0.00 & 0.15 & 0.15 & 95 & 0.03 & 0.15 & 0.14 & 92 & 0.05 & 0.15 & 0.16 & 96 & 0.02 & 0.14 & 0.14 & 95 \\ 
& 2000 & 0.00 & 0.11 & 0.10 & 94 & 0.02 & 0.11 & 0.10 & 92 & 0.03 & 0.11 & 0.11 & 95 & 0.01 & 0.10 & 0.10 & 94 \\ 
2 & 500 & 0.00 & 0.22 & 0.22 & 96 & -0.14 & 0.21 & 0.20 & 86 & -0.11 & 0.22 & 0.24 & 90 & -0.17 & 0.2 & 0.21 & 89 \\ 
 & 1000 & 0.00 & 0.15 & 0.16 & 96 & -0.04 & 0.16 & 0.14 & 90 & -0.01 & 0.17 & 0.18 & 96 & -0.07 & 0.15 & 0.15 & 92 \\ 
& 2000 & -0.01 & 0.11 & 0.11 & 94 & -0.02 & 0.12 & 0.10 & 91 & 0.01 & 0.12 & 0.13 & 95 & -0.04 & 0.11 & 0.11 & 93 \\ 
3 & 500 & 0.01 & 0.26 & 0.23 & 92 & 0.04 & 0.28 & 0.23 & 88 & -0.60 & 0.21 & 0.19 & 16 & -0.05 & 0.28 & 0.23 & 88 \\ 
& 1000 & 0.01 & 0.18 & 0.16 & 94 & 0.05 & 0.22 & 0.18 & 88 & -0.50 & 0.16 & 0.14 & 7 & 0.00 & 0.22 & 0.19 & 91 \\ 
& 2000 & 0.00 & 0.13 & 0.12 & 94 & 0.02 & 0.17 & 0.14 & 88 & -0.40 & 0.13 & 0.10 & 6 & 0.01 & 0.17 & 0.15 & 92 \\ 
4 & 500 & 0.00 & 0.22 & 0.21 & 94 & -0.10 & 0.20 & 0.19 & 90 & -0.25 & 0.21 & 0.20 & 72 & -0.09 & 0.19 & 0.19 & 93 \\ 
& 1000 & 0.00 & 0.15 & 0.15 & 94 & -0.09 & 0.15 & 0.13 & 88 & -0.24 & 0.15 & 0.14 & 57 & -0.08 & 0.14 & 0.14 & 90 \\ 
& 2000 & 0.00 & 0.11 & 0.10 & 94 & -0.04 & 0.11 & 0.10 & 87 & -0.21 & 0.11 & 0.10 & 43 & -0.05 & 0.11 & 0.10 & 90 \\ 
 \hline
\end{tabular}
}
\end{table}

\subsection{Effect modification estimands}

In a second set of simulation experiments, we considered inference for effect modification estimand (\ref{estimandintg-linear}), with $g(.)$ the identity link and without making the assumption of conditionally independent exposures. We generated a 10-dimensional covariate $L\sim N(0,\Sigma)$, 
where $\Sigma$ was randomly generated as before. The exposure was generated as in the previous section, and the outcome as $Y\sim N(3/(1+\exp(L_3-L_2))+A/(1+\exp(L_1-L_2)),1)$. This data-generating mechanism is inspired by \cite{nie2017quasi}, but made more complicated by means of a non-randomised exposure $A$. 
Our aim was to assess evidence for modification of the effect of $A$ by $L_3$. Since such effect modification is absent, we here studied the performance of different estimation methods w.r.t. their ability to retrieve zero effect modification (thus also giving us a different perspective than in the previous section, where we contrasted each estimator with its limit). The simulation results demonstrate favourable results for the proposal, based on random forests (via the `grf' package described in \cite{athey2019generalized}) as compared to OLS based on a linear model that includes all main effects along with the interaction between $A$ and $L_3$. In particular, we observe smaller bias and better coverage at the expense of a modest increase in standard errors (around 30\% larger). 

In a second set of simulation experiments, we made the data-generating mechanism even more challenging by changing the outcome model to $Y\sim N(3/(1+\exp(L_3-L_2))+A/(1+\exp(L_1-L_2))+5AL_6,1)$. The inclusion of an interaction between $A$ and $L_6$ now makes it increasingly difficult to demonstrate the absence of effect modification between $A$ and $L_3$ (which has a correlation of -0.54 with $L_6$).
The simulation results demonstrate drastically favourable results for the proposal with a much smaller bias as well as standard errors (up to 4 times smaller than for OLS), resulting in much better coverage. 

To demonstrate the behaviour under conditions where the linear regression model is correctly specified, we additionally generated a continuous exposure $A\sim N(\gamma^T L,1)$, where $\gamma$ is the $d$-dimensional unit vector scaled by $1/\sqrt{40}$, and the outcome as $Y\sim N(\gamma^T L+5AL_3,1)$. Both methods give good performance in this setting, with the proposal not surprisingly delivering larger standard errors (roughly up to 2.5 times larger). Here, better performance can be expected with the use of ensemble learners. 

\begin{table}
\caption{Simulation results on effect modification: empirical bias (Bias) and standard deviation (Emp SD), sample average of the estimated influence-curve based standard errors (Mean SE), and coverage of 95\% Wald confidence intervals (Cov).}
\centering
\begin{tabular}{llllllllll}\hline
Exper. & $n$ & \multicolumn{4}{c}{OLS} & \multicolumn{4}{c}{Proposal}\\
& & Bias & Emp SD & Mean SE & Cov & Bias & Emp SD & Mean SE & Cov\\\hline
1 & 500 & -0.047 & 0.051 & 0.051 & 84 & -0.034 & 0.067 & 0.073 & 95\\
& 1000 & -0.046 & 0.037& 0.036 & 76 &-0.016& 0.051& 0.050 & 93 \\
& 2000 & -0.046 & 0.027 & 0.025 & 55 & -0.015 & 0.036 & 0.035 & 92 \\
2 & 500 & -2.92 & 0.24 & 0.23 & 0 & -0.31 & 0.15 & 0.16 & 49\\
& 1000 & -2.92 & 0.17 & 0.16 & 0 & -0.12 & 0.077 & 0.085 & 77\\
& 2000 & -2.92 & 0.11 & 0.11 & 0 & -0.057 & 0.044 & 0.052 & 88 \\
3 & 500 & 0.00 & 0.015 & 0.015 & 94 & 0.019 & 0.042 & 0.043 & 93\\
& 1000 & 0.00 & 0.010 & 0.010 & 95 & 0.013 & 0.027 & 0.029 & 95\\
& 2000 & 0.00 & 0.007 & 0.007 & 97 & 0.002 & 0.018 & 0.021 & 97\\\hline
\end{tabular}
\end{table}

\section{Data analysis}

The First Steps program was set up in 1989 in Washington State, United States, in order to serve low-income pregnant woman and children. A specific goal was to reduce the risk of low birth weight. Using data obtained from birth certificates from 2,500 children born in King County, Washington in 2001, we sought to evaluate the effects of the First Steps program on infant birthweight, as well as its association with maternal age. We were also interested in the possible interaction between the two exposures considered. 

We first carried out a more traditional analyses using parametric models. Specifically, we fit a linear model for infant birth weight (in grams), with an indicator participation on the First Steps program and maternal age as predictors, as well other baseline covariates (child's sex, mother's age, race (asian, black, hispanic, white or other), smoking status and marital status). This model yielded estimates of -13.57 (95\% CI: -76.3, 49.2) for First Steps participation. Assuming that we have adjusted for all common causes of First Steps participation and birth weight, and additionally that the linear model is correctly specified, then the first regression coefficient suggests that participation in the program led to an average reduction of -13.57 grams in birth weight (although the confidence interval contained the null). For comparison, fitting a linear model unadjusted for covariates yielded an estimate of -66.18 (95\% CI: -125.79, -6.57), such that ignoring confounding gives the impression that the intervention was harmful. We then refit the linear model with an interaction term; it was estimated that the association between program participation and birth weight increased by 2.7 units per year increase in maternal age (95\% CI: -7.00, 12.3).  We fit a separate linear model, adjusted for all other covariates except program participation, to assess the effect of age which was estimated as 0.037 (95\% CI: -4.40, 4.47). We did not adjust for participation given that it was an externally introduced factor that may be predicted by age.

We repeated this analysis after dichotomising the outcome (an infant was considered to have low birth weight if they weighed $<2,500g$). The estimated log-odds ratios for low birth weight were -0.038 (95\% CI: -0.55, 0.44) for First Steps participation and 0.037 for age (95\% CI: 0.00, 0.073), again taken from separate models. In the larger model, the interaction was estimated as 0.022 (95\%: -0.051, 0.093). 

We re-analysed the data using the methods proposed in this article; first we estimated the propensity-overlap weighted effect of First Steps participation on birth weight using the influence curve-based estimator in (\ref{robinson_88}). The nuisance functionals $E(A|L)$ and $E(Y|L)$ (along with all others described in the section) were estimated using random forests via the `grf' package (with tuning parameters chosen using cross-validation). We obtained an estimate of -3.53 (95\% CI: -71.61, 64.55), which was smaller in magnitude than in the previous analysis, and reflects our a priori belief that program participation is unlikely to lead to a strong decrease in infant birth weight. In looking at the weighted effect of maternal age, we again did not adjust for program participation. The proposal yielded an estimate of -0.78 (95\% CI: -5.65, 4.09). By construction, these can be interpreted as the main effects of First Steps participation and age, regardless of the presence of possible interactions. In a subsequent analysis, we also re-estimated the interaction between First Steps participation and maternal age without making assumptions about possible dependencies between these exposures, and found the interaction to be more pronounced. We obtained an estimate of 7.18 (95\% CI: -5.82, 20.19). Repeating this analysis for the weighted average of log-odds of low birthweight ratios gave the effect of program participation as 0.15 (95\% CI: -0.49, 0.80) and maternal age as 0.055 (95\% CI: 0.013, 0.097). 


\section{Discussion}

We have emphasised that most data analyses rely on modelling assumptions in more intricate ways than we may realise. 
They extract information from those assumptions, rather than from the data alone. This may result in estimators for, for instance, a conditional association that are not guaranteed to summarise that association well (e.g. that cannot be viewed as a weighted average of covariate-specific conditional association measures) when those modelling assumptions fail. It may moreover deliver overly optimistic uncertainty assessments, even when based on sandwich standard errors, that are only justified when those modelling assumptions hold. With others, we therefore recommend that the starting point of a data analysis becomes the choice of an estimand, as opposed to the choice of a model. This ensures that the analysis' aim is unambiguously clear at all times, regardless of issues of model misspecification, and that uncertainty assessments, by virtue of being obtained under the nonparametric model, reflect solely the information that is contained in the data. To prevent that this is rendering interpretation more complicated, we have chosen to focus on estimands that can be interpreted as familiar regression parameters when corresponding models hold, but continue to capture what these parameters aim to summarise when these models are misspecified.

The idea of starting the analysis with the choice of an estimand, has become well integrated in causal inference research \citep{hernan2010causal}. 
Here, estimands are typically chosen with a view on specific interventions, whose impact one aims to assess. This literature has primarily focused on the average causal effect, $E\left(Y^1-Y^0\right)$, which expresses how different the expected outcome would be if all subjects in the population were treated versus untreated, and is useful - in fact, often more useful than the estimands we consider - if such interventions can be conceived. For a continuous exposure, contrasts of  $E\left(Y^a\right)$ for different exposure levels $a$ are arguably less meaningful as interventions that force each one's exposure to take on level $a$ may not be realistic (consider e.g. the effect of fixing everyone's BMI at 25) and demand enormous extrapolations. Continuous exposures moreover demand a greater need to summarise, which is naturally done by means of so-called marginal structural models \citep{robins2000marginal}, such as  
\[E(Y^a)=\alpha+\beta a,\]
for all $a$. Weighted least squares regression of $Y$ on $A$, using so-called stabilised weights $f(A)/f(A|L)$, then delivers an estimator for $\beta$ whose probability limit equals
\[\frac{E\left(\mbox{\rm Cov}\left[A^*,E\left\{E(Y|A^*,L)|L\right\}\right]\right)}{\mbox{\rm Var}\left(A\right)},\]
where $A^*$ is a random draw from the marginal distribution of $A$. This expression shows that while the starting point of a causal analysis is often an explicit estimand, also here, the desire to summarise high-dimensional information often leads one to working with estimands that are implicitly defined by the estimation procedure, as is most pronounced in studies that investigate the effects of time-varying exposures. This is undesirable. 

In causal inference applications, this explicit need for summarisation can be avoided by focussing on estimands that depend on the natural value of treatment \citep{hubbard2008population,munoz2012population,young2014identification}, for instance, that consider the effect of shifting the exposure with one unit:
\[E\left(Y^{A+1}-Y^{A}\right).\]
This estimand, which also reduces to $\beta$ in model (\ref{gplm}) with identity link when that model is correctly specified, is directly relevant if interest lies in the effect of interventions that aim to increase the exposure by one unit. In such settings, it is easier to interpret than the estimand (\ref{estimand-main}). It has the drawback, however, that such specific interventions may be rare and that the estimand is very specific to the chosen intervention. In particular, since $E\left(Y^{A+2}-Y^{A}\right)$ will not generally equal twice $E\left(Y^{A+1}-Y^{A}\right)$, a need to summarise the effects $E\left(Y^{A+a}-Y^{A}\right)/a$ for different levels of $a$ may remain when there is no convincing reason to consider $a=1$. In this paper, we have therefore opted to work with more generic estimands, that are also relevant when no specific interventions are considered (e.g. when describing the association of outcome with age, when measuring time trends, ...), and whose influence curve does not involve inverse weighting by the conditional density of $A$, given $L$. Such inverse weighting complicates the use of machine learning procedures (e.g., it may require the need for binning, as in \cite{munoz2012population}), and the need for it signals extrapolations being made (e.g., the fact that a one-unit increase in exposure may be very unlikely for subjects in certain covariate strata). In this paper, we have therefore focussed on estimands with a generic definition (regardless of whether the exposure is discrete or continuous, and regardless of whether one aims to answer a causal question or not), for which inference can be developed in a generic way (regardless of whether the exposure is discrete or continuous). Such generic estimands are important to enable broadly accessible data analyses.
Arguably, a drawback of those considered estimands is that they depend on the exposure distribution, as is for instance seen in (\ref{overlap-linear}). Such weighting may be considered undesirable (in a similar way that the partial likelihood estimator of the hazard ratio under a Cox model has been criticised for its limit depending on the censoring distribution in a complicated manner \citep{van_der_laan_targeted_2011}); however, it is the unavoidable consequence of working with estimands that avoid strong extrapolations away from the observed exposure distribution.

In our attempt to come up with generic estimands for regression parameters, we have experienced a need for clear principles for choosing estimands, as opposed to letting them be mere projection parameters \citep{buja_models_2019}. In the considered context, we have found it useful to start from the premise that $E(Y|A,L)$ is known for all levels of $A$ and $L$, and to consider how to best summarise this information when the aim is parsimony. This is best done with some regression model in mind, to ensure that the estimand coincides with a familiar regression parameter when that model is correctly specified, and thus remains well interpretable. To prevent that the assumptions embodied in the entire regression model dominate the choice of estimand, we have focussed on (generalised) partially linear models, which merely specify the conditional association or effect modification term of interest. The population limit of semiparametric estimators under such model may then serve as a template for a choice of estimand. Such choice is non-unique. In our work, we have aimed for simplicity, realising that other estimands (e.g. that involve inverse weighting by the conditional outcome variance) can be inferred more efficiently in the presence of heteroscedasticity. Our choice was further guided by the desire to prevent inverse weighting by the exposure density for the aforementioned reasons. This turned out non-trivial in the case of effect modification estimands. In future work, we hope to make similar developments for parameters indexing proportional hazard models for time-to-event data and marginal models for repeated measures data.

\section*{Acknowledgement}
The authors would like to thank Mark van der Laan for inspiring discussions that have influenced this work, and 
gratefully acknowledge support from BOF Grant BOF.01P08419. 

\bibliographystyle{rss}
\bibliography{EstimandsGLMref}

\section*{Appendix A: Calculation of the influence curve of (\ref{estimand-main})}

We first calculate the efficient influence curve of 
\begin{align*}
\theta(\beta)&=E\left(\left\{A^*-E(A|L)\right\}\left[g\{E(Y|A^*,L)\}-\beta A^*)\right]\right)\\&=\int \left\{A-E(A|L)\right\}\left[g\{E(Y|A,L)\}-\beta A\right]f(A,L)dAdL
\end{align*}
under the nonparametric model for the observed data $O=(Y,A,L)$. Taking the derivative w.r.t. the scalar parameter $t$ indexing a one-dimensional parametric submodel of $f(O)$ (which returns $f(O)$ at $t=0$),
 we find that 
\begin{eqnarray*}
\frac{\partial \theta(\beta)}{\partial t}|_{t=0}&=&\int \left\{A-E(A|L)\right\}g'\{E(Y|A,L)\}YS_t(Y|A,L)f(O)dO\\
&&+\int \left\{A-E(A|L)\right\}\left\{g\{E(Y|A,L)\}-\beta A\right\}S_t(A|L)f(O)dO\\
&&+\int \left\{A-E(A|L)\right\}\left[g\{E(Y|A,L)\}-\beta A\right]S_t(L)f(O)dO\\
&&-\int E\left[g\{E(Y|A,L)\}-\beta A|L\right]AS_t(A|L)f(O)dO
\end{eqnarray*}
where $S_t(Y|A,L),S_t(A|L)$ and $S_t(L)$ are the scores w.r.t. $t$ in that parametric submodel, corresponding to the distributions $f(Y|A,L),f(A|L)$ and $f(L)$, respectively. With $S_t(O)=S_t(Y|A,L)+S_t(A|L)+S_t(L)$, it follows by the mean zero property of scores and the fact that $\theta(\beta)=0$ that
\begin{eqnarray*}
\frac{\partial \theta(\beta)}{\partial t}|_{t=0}&=&\int \left\{A-E(A|L)\right\}g'\{E(Y|A,L)\}\left\{Y-E\left(Y|A,L\right)\right\}S_t(O)f(O)dO\\
&&+\int \left\{A-E(A|L)\right\}\left[g\{E(Y|A,L)\}-\beta A\right]S_t(O)f(O)dO\\
&&-\int E\left(\left\{A-E(A|L)\right\}\left[g\{E(Y|A,L)\}-\beta A\right]|L\right)S_t(O)f(O)dO\\
&&+\int E\left(\left\{A-E(A|L)\right\}\left[g\{E(Y|A,L)\}-\beta A\right]|L\right)S_t(O)f(O)dO\\
&&-E\left(\left\{A-E(A|L)\right\}\left[g\{E(Y|A,L)\}-\beta A\right]\right)\int S_t(O)f(O)dO\\
&&-\int E\left[g\{E(Y|A,L)\}-\beta A|L\right]\left\{A-E(A|L)\right\}S_t(O)f(O)dO\\
&=&\int \left\{A-E(A|L)\right\}\left[\mu(Y,A,L)-\beta\{A- E(A|L)\} \right]S_t(O)f(O)dO.
\end{eqnarray*}
It now follows from Theorem 2.2 in \cite{newey1990semiparametric} that the efficient influence curve of $\theta(\beta)$ under the nonparametric model equals
\[\left\{A-E(A|L)\right\}\left[\mu(Y,A,L)-\beta \left\{A-E(A|L)\right\}\right].\]
It further follows from
\[\frac{\partial \theta(\beta)}{\partial t}=\frac{\partial \theta(\beta)}{\partial \beta}\frac{\partial \beta}{\partial t}\]
that the efficient influence curve for $\beta$ is 
\[-\frac{\left\{A-E(A|L)\right\}\left[\mu(Y,A,L)-\beta \left\{A-E(A|L)\right\}\right]}{E\left[\left\{A-E(A|L)\right\}^2\right]}.\]

In \cite{crump2006moving}, the efficient influence curve of the parameter (\ref{estimand-main}) is given as 
\begin{align*}
&\left(E\left[\pi(L)\{1-\pi(L)\}\right]\right)^{-1}\\
&\times \big[\{1-\pi(L)\}A\{Y-E(Y|A=1,L)\}-\pi(L)(1-A)\{Y-E(Y|A=0,L)\}\\
& +  \pi(L)\{1-\pi(L)\}\{E(Y|A=1,L)-E(Y|A=0,L)-\beta\}\\
& + \{A-\pi(L)\}\{1-2\pi(L)\}\{E(Y|A=1,L)-E(Y|A=0,L)-\beta\}\big].
\end{align*}
The numerator of the expression can be rearranged as
\begin{align*}
& \{A-\pi(L)\}Y-\{1-\pi(L)\}AE(Y|A=1,L)+\pi(L)(1-A)E(Y|A=0,L)\\
&+\{A-\pi(L)\}^2\{E(Y|A=1,L)-E(Y|A=0,L)-\beta\}\\
&=\{A-\pi(L)\}[Y-\pi(L)E(Y|A=1,L)-\{1-\pi(L)\}E(Y|A=0,L)-\beta\{A-\pi(L)\}]\\
&=\{A-\pi(L)\}[Y-E(Y|L)-\beta\{A-\pi(L)\}]
\end{align*}
and therefore our influence function coincides with theirs when $A$ is binary and $g(\cdot)$ is the identity link function. 

Using a Von Mises expansion (see e.g. \citet{van2000asymptotic}) we have that 
\begin{align*}
\sqrt{n}(\hat{\beta}-\beta)=&\sqrt{n}\hat{E}\left(\frac{\left\{A-E(A|L)\right\}\left[\mu(Y,A,L)-\beta \left\{A-E(A|L)\right\}\right]}{E\left[\left\{A-E(A|L)\right\}^2\right]}\right)+R_1+R_2\end{align*}
where $\hat{E}(.)$ refers to the sample average,
\begin{align*}
R_1=\sqrt{n}(\hat{E}-E)&\left(\frac{\left\{A-\hat{E}(A|L)\right\}\left[\hat{\mu}(Y,A,L)-\hat{\beta} \left\{A-\hat{E}(A|L)\right\}\right]}{\hat{E}\left[\left\{A-\hat{E}(A|L)\right\}^2\right]}\right.\\
&\left.-\frac{\left\{A-E(A|L)\right\}\left[\mu(Y,A,L)-\beta \left\{A-E(A|L)\right\}\right]}{E\left[\left\{A-E(A|L)\right\}^2\right]}\right)
\end{align*}
and
\begin{align*}
R_2=\sqrt{n}(\hat{\beta}-\beta)+\sqrt{n}E\left(\frac{\left\{A-\hat{E}(A|L)\right\}\left[\hat{\mu}(Y,A,L)-\hat{\beta} \left\{A-\hat{E}(A|L)\right\}\right]}{\hat{E}\left[\left\{A-\hat{E}(A|L)\right\}^2\right]}\right).
\end{align*}
Throughout this section, for a function $f(O)$ of the data $O$ we use the notation $E\{f(O)\}=\int f(O) P(O)dO$; for an estimate $\hat{f}$, $E\{\hat{f}(O)\}$ averages over $O$ but not $\hat{f}$. The term $R_1$ can typically be shown to be $o_p(1)$ using either empirical process conditions or sample-splitting. In what follows, we aim to derive $R_2$ and understand under what conditions it is asymptotically negligible.

From the definition of $R_2$, it follows that 
\begin{eqnarray*}
R_2=&&\sqrt{n}(\hat{\beta}-\beta)-\sqrt{n}\hat{\beta}\frac{E[\{A-\hat{E}(A|L)\}^2]}{\hat{E}[\{A-\hat{E}(A|L)\}^2]}\\
&&+\frac{\sqrt{n}E\left\{\{A-\hat{E}(A|L)\}\left(g\{\hat{E}(Y|A,L)\}-\hat{E}[g\{\hat{E}(Y|A,L)\}|L]\right)
\right\}}{\hat{E}[\{A-\hat{E}(A|L)\}^2]}\\
&&+\frac{\sqrt{n}E\left(A-\hat{E}(A|L)\left[g'\{\hat{E}(Y|A,L)\}\{Y-\hat{E}(Y|A,L)\}\right]\right)}
{\hat{E}[\{A-\hat{E}(A|L)\}^2]}\\
&&=\sqrt{n}(\hat{\beta}-\beta)\left[1-\frac{E[\{A-\hat{E}(A|L)\}^2]}{\hat{E}[\{A-\hat{E}(A|L)\}^2]}\right]\\
&&+\frac{\sqrt{n}E\left\{\{A-\hat{E}(A|L)\}\left(g\{\hat{E}(Y|A,L)\}-\hat{E}[g\{\hat{E}(Y|A,L)\}|L]-\beta \{A-\hat{E}(A|L)\}\right)\right\}}{\hat{E}[\{A-\hat{E}(A|L)\}^2]}\\
&&+\frac{\sqrt{n}E\left(\{A-\hat{E}(A|L)\}\left[g'\{\hat{E}(Y|A,L)\}\{Y-\hat{E}(Y|A,L)\}\right]\right)}
{\hat{E}[\{A-\hat{E}(A|L)\}^2]},
\end{eqnarray*}
Here, the first term is $o_p(|\sqrt{n}(\hat{\beta}-\beta)|)$, and thus a lower order term. 
To understand the behaviour of the remaining two terms, we 
use that 
\[\beta = \frac{E\left\{\{A-E(A|L)\}\left(g'\{E(Y|A,L)\}\{Y-E(Y|A,L)\}+g\{E(Y|A,L)\}-E\left[g\{E(Y|A,L)\}|L\right]\right)\right\}}{E\left[\{A-E(A|L)\}^2\right]}\]
by virtue of the estimand's definition,
and
simplify notation as follows. Let $\tau \equiv E(Y|A,L)$; then the remaining two terms can be written as 
\begin{eqnarray*}
&&\frac{\sqrt{n}}{\hat{E}[\{A-\hat{E}(A|L)\}^2]}\left\{
E\left(\left\{A-\hat{E}(A|L)\right\}\left[g'(\hat{\tau})(Y-\hat{\tau})+g(\hat{\tau})-\hat{E}\{g(\hat{\tau})|L\}\right]\right)\right.\\
&&\left.-E\left(\left\{A-E(A|L)\right\}\left[g'({\tau})(Y-{\tau})+g({\tau})-E\{g(\tau)|L\}\right]\right)\frac{E\left[\left\{A-\hat{E}(A|L)\right\}^2\right]}{{E}\left[\left\{A-E(A|L)\right\}^2\right]}\right\}\\
\end{eqnarray*}
Next, note that 
\begin{eqnarray*}
&&E\left(\{A-\hat{E}(A|L)\}\left[g'(\hat{\tau})(Y-\hat{\tau})+g(\hat{\tau})-\hat{E}\{g(\hat{\tau})|L\}\right]\right)\\
&&=E\left(\{A-E(A|L)+E(A|L)-\hat{E}(A|L)\}\left[g(\tau)+O_p\left\{(\tau-\hat{\tau})^2\right\}-\hat{E}\{g(\hat{\tau})|L\}\right]\right)
\end{eqnarray*}
and that
\[E\left(\{A-\hat{E}(A|L)\}\left[g'({\tau})(Y-{\tau})+g({\tau})-E\{g(\tau)|L\}\right]\right)=E\left(\left\{A-E(A|L)\right\}\left[g({\tau})-E\{g(\tau)|L\}\right]\right).\]
This reduces these remaining 2 terms to 
\begin{eqnarray*}
&&\frac{\sqrt{n}}{\hat{E}[\{A-\hat{E}(A|L)\}^2]}\left\{
E\left(\left\{A-E(A|L)\right\}\left[g({\tau})+O_p\left\{(\tau-\hat{\tau})^2\right\}-\hat{E}\left\{g(\hat{\tau})|L\right\}\right]\right)\right.\\
&&\left.+E\left(\left\{E(A|L)-\hat{E}(A|L)\right\}\left[g({\tau})+O_p\left\{(\tau-\hat{\tau})^2\right\}-\hat{E}\left\{g(\hat{\tau})|L\right\}\right]\right)\right.\\
&&\left.-E\left(\left\{A-E(A|L)\right\}\left[g(\tau)-E \left\{g(\tau)|L\right\}\right]\right)\right.\\
&&\left.
+\left(1-\frac{E\left[\left\{A-\hat{E}(A|L)\right\}^2\right]}{{E}\left[\left\{A-E(A|L)\right\}^2\right]}\right)E\left(\left\{A-E(A|L)\right\}\left[g(\tau)-E \left\{g(\tau)|L\right\}\right]\right)\right\}\\
&&=\frac{\sqrt{n}}{\hat{E}[\{A-E(A|L)\}^2]}\left\{
E\left(\left\{A-E(A|L)\right\}\left[O_p\left\{(\tau-\hat{\tau})^2\right\}+E \left\{g(\tau)|L\right\}-\hat{E}\left\{g(\hat{\tau})|L\right\}\right]\right)\right.\\
&&\left. + E\left(\left\{E(A|L)-\hat{E}(A|L)\right\}\left[g({\tau})-{E}\left\{g({\tau})|L\right\}+O_p\left\{(\tau-\hat{\tau})^2\right\}\right]\right)\right.\\
&&\left. + E\left(\left\{E(A|L)-\hat{E}(A|L)\right\}\left[{E}\left\{g({\tau})|L\right\}-\hat{E}\left\{g({\tau})|L\right\}+\hat{E}\{g({\tau})|L\}-\hat{E}\{g(\hat{\tau})|L\}\right]\right)\right.\\
&&\left.+\left(1-\frac{E\left[\left\{A-E(A|L)+E(A|L)-\hat{E}(A|L)\right\}^2\right]}{{E}\left[\left\{A-E(A|L)\right\}^2\right]}\right)E\left(\left\{A-E(A|L)\right\}\left[g(\tau)-E \left\{g(\tau)|L\right\}\right]\right)\right\}\\
&&=\frac{\sqrt{n}}{\hat{E}[\{A-\hat{E}(A|L)\}^2]}\left\{
E\left(\left\{A-\hat{E}(A|L)\right\}O_p\left\{(\tau-\hat{\tau})^2\right\}\right)\right.\\
&&\left. + E\left(\left\{E(A|L)-\hat{E}(A|L)\right\}\left[E\left\{g({\tau})|L\right\}-\hat{E}\left\{g({\tau})|L\right\}+\hat{E}\{g({\tau})|L\}-\hat{E}\{g(\hat{\tau})|L\}\right]\right)\right.\\
&&\left.+\left(\frac{-E\left[2\left\{A-E(A|L)\right\}\left\{E(A|L)-\hat{E}(A|L)\right\}+\left\{E(A|L)-\hat{E}(A|L)\right\}^2\right]}{{E}\left[\left\{A-E(A|L)\right\}^2\right]}\right)\right.\\&&\left.\times E\left(\left\{A-E(A|L)\right\}\left[g(\tau)-E \left\{g(\tau)|L\right\}\right]\right)\right\}\\
&&=\frac{\sqrt{n}}{\hat{E}[\{A-\hat{E}(A|L)\}^2]}\left\{
E\left(\left\{A-\hat{E}(A|L)\right\}O_p\left\{(\tau-\hat{\tau})^2\right\}\right)\right.\\
&&\left. + E\left(\left\{E(A|L)-\hat{E}(A|L)\right\}\left[E\left\{g({\tau})|L\right\}-\hat{E}\left\{g({\tau})|L\right\}+\hat{E}\{g({\tau})|L\}-\hat{E}\{g(\hat{\tau})|L\}\right]\right)\right.\\
&&\left.+\frac{E\left[\left\{E(A|L)-\hat{E}(A|L)\right\}^2\right]}{{E}\left[\left\{A-E(A|L)\right\}^2\right]}E\left(\left\{A-E(A|L)\right\}\left[g(\tau)-E \left\{g(\tau)|L\right\}\right]\right)\right\},
\end{eqnarray*}
where we use that 
\begin{align*}
E\left(\left\{A-E(A|L)\right\}\left[E\{g({\tau})|L\}-\hat{E}\left\{g(\hat{\tau})|L\right\}\right]\right)&=0\\
E\left(\left\{E(A|L)-\hat{E}(A|L)\right\}\left[g({\tau})-{E}\left\{g({\tau})|L\right\}\right]\right)&=0
\end{align*} 
using the law of iterated expectation. It now follows by the Cauchy-Schwarz inequality that the remainder term converges to zero in probability under the following, fairly weak conditions
\begin{eqnarray*}
E\left(\left\{A-\hat{E}(A|L)\right\}O_p\left\{(\tau-\hat{\tau})^2\right\}\right)&=&o_p(n^{-1/2})\\
E\left(\left\{E(A|L)-\hat{E}(A|L)\right\}^2\right)^{1/2}E\left(\left[E\{g({\tau})|L\}-\hat{E}\left\{g({\tau})|L\right\}\right]^2\right)^{1/2}&=&o_p(n^{-1/2})\\
E\left(\left\{E(A|L)-\hat{E}(A|L)\right\}^2\right)^{1/2}E\left(\left[\hat{E}\left\{g({\tau})|L\right\}-\hat{E}\left\{g(\hat{\tau})|L\right\}\right]^2\right)^{1/2}&=&o_p(n^{-1/2})\\
E\left[\left\{E(A|L)-\hat{E}(A|L)\right\}^2\right]&=&o_p(n^{-1/2}),
\end{eqnarray*}
where the first condition is redundant when $g(.)$ is the identity link.

\section*{Appendix B: Calculation of the influence curve of (\ref{estimandint-linear}) when $A_1\cip A_2|L$}

We first calculate the efficient influence curve of 
\begin{eqnarray*}
\theta(\beta)&=&E\left(\left\{A_1^*-E(A_1|L)\right\}\left\{A_2^*-E(A_2|L)\right\}\left[g\{E(Y|A^*_1,A_2^*,L)\}-\beta A_1^*A_2^*\right]\right)\\
&=&\int \left\{A_1-E(A_1|L)\right\}\left\{A_2-E(A_2|L)\right\}\left[g\{E(Y|A_1,A_2,L)\}-\beta A_1A_2\right]\\&&\times\frac{f(A_1|L)f(A_2|L)}{f(A_1,A_2|L)}f(A_1,A_2,L)dA_1dA_2dL,
\end{eqnarray*}
under the nonparametric model for the observed data $O=(Y,A_1,A_2,L)$. When $A_1\cip A_2|L$, then this reduces to 
\begin{eqnarray*}
\theta(\beta)&=&\int \left\{A_1-E(A_1|L)\right\}\left\{A_2-E(A_2|L)\right\}\left[g\{E(Y|A_1,A_2,L)\}-\beta A_1A_2\right]\\
&&\times f(A_1,A_2,L)dA_1dA_2dL.
\end{eqnarray*}
In what follows, we will only rely on this assumption to define the estimand, but nonetheless infer its influence curve under the nonparametric model (in the sense that we will study paths along all parametric submodels, including those where $A_1$ and $A_2$ are conditionally dependent, given $L$).

Taking the derivative w.r.t. the scalar parameter $t$ indexing a one-dimensional parametric submodel of $f(O)$ (which returns $f(O)$ at $t=0$), we find that 
\begin{eqnarray*}
&&\frac{\partial \theta(\beta)}{\partial t}|_{t=0}\\&&=\int  \left\{A_1-E(A_1|L)\right\}\left\{A_2-E(A_2|L)\right\}g'\{E(Y|A_1,A_2,L)\}YS_t(Y|A_1,A_2,L)f(O)dO\\
&&+\int \left\{A_1-E(A_1|L)\right\}\left\{A_2-E(A_2|L)\right\}\left[g\{E(Y|A_1,A_2,L)\}-\beta A_1A_2\right]S_t(A_1,A_2|L)f(O)dO\\
&&+\int \left\{A_1-E(A_1|L)\right\}\left\{A_2-E(A_2|L)\right\}\left[g\{E(Y|A_1,A_2,L)\}-\beta A_1A_2\right]S_t(L)f(O)dO\\
&&-\int E\left(\left\{A_2-E(A_2|L)\right\}\left[g\{E(Y|A_1,A_2)\}-\beta A_1A_2\right]|L\right)A_1S_t(A_1,A_2|L)f(O)dO\\
&&-\int E\left(\left\{A_1-E(A_1|L)\right\}\left[g\{E(Y|A_1,A_2)\}-\beta A_1A_2\right]|L\right)A_2S_t(A_1,A_2|L)f(O)dO.
\end{eqnarray*}
where $S_t(Y|A_1,A_2,L),S_t(A_1,A_2|L)$ and $S_t(L)$ are the scores w.r.t. $t$ in that parametric submodel, corresponding to the distributions $f(Y|A_1,A_2,L),f(A_1,A_2|L)$ and $f(L)$, respectively.
With $S_t(O)=S_t(Y|A_1,A_2,L)+S_t(A_1,A_2|L)+S_t(L)$, it follows by the mean zero property of scores and the fact that $\theta(\beta)=0$ that
\begin{eqnarray*}
&&\frac{\partial \theta(\beta)}{\partial t}|_{t=0}\\&&=\int  \left\{A_1-E(A_1|L)\right\}\left\{A_2-E(A_2|L)\right\}g'\{E(Y|A_1,A_2,L)\}\left\{Y-E(Y|A_1,A_2,L)\right\}S_t(O)f(O)dO\\
&&+\int \left\{A_1-E(A_1|L)\right\}\left\{A_2-E(A_2|L)\right\}\left[g\{E(Y|A_1,A_2,L)\}-\beta A_1A_2\right]S_t(O)f(O)dO\\
&&-\int E\left(\left\{A_1-E(A_1|L)\right\}\left\{A_2-E(A_2|L)\right\}\left[g\{E(Y|A_1,A_2,L)\}-\beta A_1A_2\right]|L\right)S_t(O)f(O)dO\\
&&+\int E\left(\left\{A_1-E(A_1|L)\right\}\left\{A_2-E(A_2|L)\right\}\left[g\{E(Y|A_1,A_2,L)\}-\beta A_1A_2\right]|L\right)S_t(O)f(O)dO\\
&&-\int E\left(\left\{A_2-E(A_2|L)\right\}\left[g\{E(Y|A_1,A_2,L)\}-\beta A_1A_2\right]|L\right) \left\{A_1-E(A_1|L)\right\}S_t(O)f(O)dO\\
&&-\int E\left(\left\{A_1-E(A_1|L)\right\}\left[g\{E(Y|A_1,A_2,L)\}-\beta A_1A_2\right]|L\right) \left\{A_2-E(A_2|L)\right\}S_t(O)f(O)dO\\
&&=\int  \left\{A_1-E(A_1|L)\right\}\left\{A_2-E(A_2|L)\right\}g'\{E(Y|A_1,A_2,L)\}\left\{Y-E(Y|A_1,A_2,L)\right\}S_t(O)f(O)dO\\
&&+\int \left\{A_1-E(A_1|L)\right\}\left\{A_2-E(A_2|L)\right\}\left[g\{E(Y|A_1,A_2,L)\}-\beta A_1A_2\right]S_t(O)f(O)dO\\
&&-\int E\left(\left\{A_2-E(A_2|L)\right\}\left[g\{E(Y|A_1,A_2,L)\}-\beta A_1A_2\right]|L\right) \left\{A_1-E(A_1|L)\right\}S_t(O)f(O)dO\\
&&-\int E\left(\left\{A_1-E(A_1|L)\right\}\left[g\{E(Y|A_1,A_2,L)\}-\beta A_1A_2\right]|L\right) \left\{A_2-E(A_2|L)\right\}S_t(O)f(O)dO
\end{eqnarray*}
Using a similar argument as in the previous appendix, we conclude that the influence curve for $\beta$ equals
\begin{eqnarray*}
&&-\left(E\left[\left\{A_1-E(A_1|L)\right\}\left\{A_2-E(A_2|L)\right\}A_1A_2\right]\right)^{-1}\\
&&\times \left(\left\{A_1-E(A_1|L)\right\}\left\{A_2-E(A_2|L)\right\}g'\{E(Y|A_1,A_2,L)\}\left\{Y-E(Y|A_1,A_2,L)\right\}\right.\\
&&\left. + \left\{A_1-E(A_1|L)\right\}\left\{A_2-E(A_2|L)\right\} \left[g\{E(Y|A_1,A_2,L)\}-\beta A_1A_2\right]\right. \\
&&\left. - E\left(\left\{A_2-E(A_2|L)\right\}\left[g\{E(Y|A_1,A_2)\}-\beta A_1A_2\right]|L\right) \left\{A_1-E(A_1|L)\right\}\right.\\
&&\left. - E\left(\left\{A_1-E(A_1|L)\right\}\left[g\{E(Y|A_1,A_2)\}-\beta A_1A_2\right]|L\right) \left\{A_2-E(A_2|L)\right\}\right).
\end{eqnarray*}
That the remainder term in the asymptotic expansion of $\hat{\beta}$ converges to zero in probability follows as a special case of the proof given in the next section.

\section*{Appendix C: Calculation of the influence curve of (\ref{estimandintg-linear})}

Define 
\[\Lambda=\left\{d_1(A_1,L)+d_2(A_2,L): d_1(.), d_2(.) \in L_2(P) \ \mbox{\rm arbitrary}\right\},\]
whose orthocomplement in the Hilbert space of functions of $(A_1,A_2,L)$ in $L_2(P)$, equipped with the covariance as inner product, equals
\[\Lambda^{\perp}=\left\{d(A_1,A_2,L)\in L_2(P): E\left\{d(A_1,A_2,L)|A_1,L\right\}=E\left\{d(A_1,A_2,L)|A_2,L\right\}=0\right\}\]
(Vansteelandt et al., 2008). Let $P(.)$ be the orthogonal projection operator onto $\Lambda^{\perp}$.
Then our focus is on the estimand defined by the solution to 
\[\theta(\beta)=E\left(P(A_1A_2)\left[g\{E(Y|A_1,A_2,L)\}-\beta A_1A_2\right]\right)=0.\]
As in previous sections, we first calculate the influence curve of $\theta(\beta)$
under the nonparametric model for the observed data $O=(Y,A,L)$. Taking the derivative w.r.t. the scalar parameter $t$ indexing a one-dimensional parametric submodel of $f(O)$ (which returns $f(O)$ at $t=0$),
 we find that 
\begin{eqnarray*}
\frac{\partial \theta(\beta)}{\partial t}|_{t=0}&=&\int P(A_1A_2)g'\{E(Y|A_1,A_2,L)\}\left\{Y-E(Y|A_1,A_2,L)\right\}S_t(Y|A,L)f(O)dO\\
&&+\int P(A_1A_2)\left[g\{E(Y|A_1,A_2,L)\}-\beta A_1A_2\right]S_t(A|L)f(O)dO\\
&&+\int P(A_1A_2)\left[g\{E(Y|A_1,A_2,L)\}-\beta A_1A_2\right]S_t(L)f(O)dO\\
&&+\int \frac{\partial P(A_1A_2)}{\partial t}|_{t=0}\left[g\{E(Y|A_1,A_2,L)\}-\beta A_1A_2\right]f(O)dO,
\end{eqnarray*}
where $S_t(Y|A,L),S_t(A|L)$ and $S_t(L)$ are the scores w.r.t. $t$ in that parametric submodel, corresponding to the distributions $f(Y|A,L),f(A|L)$ and $f(L)$, respectively. 
Since $P(A_1A_2)$ equals $A_1A_2$ minus its projection onto $\Lambda$, and since the derivative of $A_1A_2$ w.r.t. $t$ equals zero and $\Lambda$ is closed under the derivative, we have that 
\[\frac{\partial P(A_1A_2)}{\partial t}|_{t=0}\in \Lambda.\]
Since for $j=1,2$
\[E\left\{P(A_1A_2)|A_j,L\right\}=0\]
we further have that for $j=1,2$
\begin{eqnarray*}
0&=&\int \frac{\partial P(A_1A_2)}{\partial t}|_{t=0}\frac{f(A_1,A_2|L)}{f(A_j|L)}dA_{-j}+\int P(A_1A_2)S_t(A_1,A_2|L)\frac{f(A_1,A_2|L)}{f(A_j|L)}dA_{-j}\\
&&-\int P(A_1A_2)\frac{f(A_1,A_2|L)}{f(A_j|L)^2}\left\{\int S_t(A_1,A_2|L)f(A_1,A_2|L)dA_{-j}\right\}dA_{-j}\\
&=&E\left\{\frac{\partial P(A_1A_2)}{\partial t}|_{t=0}+P(A_1A_2)S_t(A_1,A_2|L)|A_j,L\right\}\\&&-E\left\{P(A_1A_2)|A_j,L\right\}E\left\{S_t(A_1,A_2|L)|A_j,L\right\}\\
&=&E\left\{\frac{\partial P(A_1A_2)}{\partial t}|_{t=0}+P(A_1A_2)S_t(A_1,A_2|L)|A_j,L\right\},
\end{eqnarray*}
so that 
\[\frac{\partial P(A_1A_2)}{\partial t}|_{t=0}+P(A_1A_2)S_t(A_1,A_2|L)\in \Lambda^{\perp}.\]
It follows that 
\begin{eqnarray*}
&&\int \left\{\frac{\partial P(A_1A_2)}{\partial t}|_{t=0}+P(A_1A_2)S_t(A_1,A_2|L)\right\}\left[g\{E(Y|A_1,A_2,L)\}-\beta A_1A_2\right]f(O)dO\\
&&=\int \left\{\frac{\partial P(A_1A_2)}{\partial t}|_{t=0}+P(A_1A_2)S_t(A_1,A_2|L)\right\}P\left[g\{E(Y|A_1,A_2,L)\}-\beta A_1A_2\right]f(O)dO\\
&&=\int P(A_1A_2)S_t(A_1,A_2|L)P\left[g\{E(Y|A_1,A_2,L)\}-\beta A_1A_2\right]f(O)dO.
\end{eqnarray*}
With $S_t(O)=S_t(Y|A,L)+S_t(A|L)+S_t(L)$, it now follows by the mean zero property of scores and the fact that $\theta(\beta)=0$ that
\begin{eqnarray*}
\frac{\partial \theta(\beta)}{\partial t}|_{t=0}&=&\int P(A_1A_2)g'\{E(Y|A_1,A_2,L)\}\left\{Y-E(Y|A_1,A_2,L)\right\}S_t(O)f(O)dO\\
&&+\int \left\{P(A_1A_2)P\left[g\{E(Y|A_1,A_2,L)\}-\beta A_1A_2\right]\right.\\
&&\left.-E\left(P(A_1A_2)P\left[g\{E(Y|A_1,A_2,L)\}-\beta A_1A_2\right]|L\right)\right\}S_t(O)f(O)dO\\
&&+\int E\left(P(A_1A_2)\left[g\{E(Y|A_1,A_2,L)\}-\beta A_1A_2\right]|L\right)S_t(O)f(O)dO.
\end{eqnarray*}
We conclude that the influence curve for $\theta(\beta)$ equals
\begin{eqnarray*}
&&P(A_1A_2)\left(g'\{E(Y|A_1,A_2,L)\}\{Y-E(Y|A_1,A_2,L)\}+P\left[g\{E(Y|A_1,A_2,L)\}-\beta A_1A_2\right]\right)\\
&&+E\left\{P(A_1A_2)
\left(\left[g\{E(Y|A_1,A_2,L)\}-\beta A_1A_2\right]-P\left[g\{E(Y|A_1,A_2,L)\}-\beta A_1A_2\right]\right)|L\right\} \\
&&=P(A_1A_2)\left(g'\{E(Y|A_1,A_2,L)\}\{Y-E(Y|A_1,A_2,L)\}+P\left[g\{E(Y|A_1,A_2,L)\}-\beta A_1A_2\right]\right)
\end{eqnarray*}
since $g\{E(Y|A_1,A_2,L)\}-\beta A_1A_2-P\left[g\{E(Y|A_1,A_2,L)\}-\beta A_1A_2\right]\in \Lambda$. Using similar arguments as before, it follows that the influence curve for $\beta$ equals
\[\frac{P(A_1A_2)}{E\left\{P(A_1A_2)^2\right\}}\left(g'\{E(Y|A_1,A_2,L)\}\{Y-E(Y|A_1,A_2,L)\}+P\left[g\{E(Y|A_1,A_2,L)\}-\beta A_1A_2\right]\right)\]

To study the conditions under which the proposed estimator is asymptotically linear with influence function given by the above influence curve, we derive the remainder term in the asymptotic expansion of $\hat{\beta}$, which is given by 
\begin{eqnarray*}
&&\sqrt{n}(\hat{\beta}-\beta)-\sqrt{n}\hat{\beta}\frac{E\left\{\hat{P}(A_1A_2)^2\right\}}{\hat{E}\left\{\hat{P}(A_1A_2)^2\right\}}+\sqrt{n}\frac{E\left(\hat{P}(A_1A_2)\hat{P}\left[g\{\hat{E}(Y|A_1,A_2,L)\}\right]\right)}{\hat{E}\left\{\hat{P}(A_1A_2)^2\right\}}\\
&&+\frac{E\left(\hat{P}(A_1A_2)\left[g'\{\hat{E}(Y|A_1,A_2,L)\}\{Y-\hat{E}(Y|A_1,A_2,L)\}\right]\right)}
{\hat{E}\left\{\hat{P}(A_1A_2)^2\right\}}\\
&&=\sqrt{n}(\hat{\beta}-\beta)\left[1-\frac{E\left\{\hat{P}(A_1A_2)^2\right\}}{\hat{E}\left\{\hat{P}(A_1A_2)^2\right\}}\right]+\sqrt{n}\frac{E\left(\hat{P}(A_1A_2)\hat{P}\left[g\{\hat{E}(Y|A_1,A_2,L)\}-\beta A_1A_2\right]\right)}{\hat{E}\left\{\hat{P}(A_1A_2)^2\right\}}\\
&&+\frac{E\left(\hat{P}(A_1A_2)\left[g'\{\hat{E}(Y|A_1,A_2,L)\}\{Y-\hat{E}(Y|A_1,A_2,L)\}\right]\right)}
{\hat{E}\left\{\hat{P}(A_1A_2)^2\right\}},
\end{eqnarray*}
As before, the first term is $o_p(|\sqrt{n}(\hat{\beta}-\beta)|)$. 
We will use that 
\[\beta = \frac{E\left\{P(A_1A_2)\left(g'\{E(Y|A_1,A_2,L)\}\{Y-E(Y|A_1,A_2,L)\}+P\left[g\{E(Y|A_1,A_2,L)\}\right]\right)\right\}}{E\left\{P(A_1A_2)^2\right\}}\]
by virtue of the estimand's definition,
and
simplify notation as follows. Let $\mu\equiv E(Y|A_1,A_2,L)$ and rewrite $A_1A_2$ as $V$. Further, rewrite the orthogonal projection $P(V)$ and $\hat{P}(V)$ for a random variable $V$ as $V-\Pi(V)$ and $V-\hat{\Pi}(V),$ respectively, where $\Pi(V),\hat{\Pi}(V)\in \Lambda$, and likewise rewrite $P\left[g\{{E}(Y|A_1,A_2,L)\}\right]$ as $g(\mu)-\Pi \left\{g(\mu)\right\}$. 
Using the same arguments as in the previous appendices, the remaining two terms can be written as 
\begin{eqnarray*}
&&\frac{\sqrt{n}}{\hat{E}\left[\left\{D-\hat{\Pi}(D)\right\}^2\right]}\left\{
E\left(\left\{D-\Pi(D)+\Pi(D)-\hat{\Pi}(D)\right\}\left[g({\mu})+O_p\left\{(\mu-\hat{\mu})^2\right\}-\hat{\Pi}\left\{g(\hat{\mu})\right\}\right]\right)\right.\\
&&\left.-E\left(\left\{D-\Pi(D)\right\}\left[g(\mu)-\Pi \left\{g(\mu)\right\}\right]\right)\right.\\
&&\left.
+\left(1-\frac{E\left[\left\{D-\hat{\Pi}(D)\right\}^2\right]}{{E}\left[\left\{D-{\Pi}(D)\right\}^2\right]}\right)E\left(\left\{D-\Pi(D)\right\}\left[g(\mu)-\Pi \left\{g(\mu)\right\}\right]\right)\right\}\\
&&=\frac{\sqrt{n}}{\hat{E}\left[\left\{D-\hat{\Pi}(D)\right\}^2\right]}\left\{
E\left(\left\{D-\Pi(D)\right\}\left[O_p\left\{(\mu-\hat{\mu})^2\right\}+\Pi \left\{g(\mu)\right\}-\hat{\Pi}\left\{g(\hat{\mu})\right\}\right]\right)\right.\\
&&\left. + E\left(\left\{\Pi(D)-\hat{\Pi}(D)\right\}\left[g({\mu})-{\Pi}\left\{g({\mu})\right\}+O_p\left\{(\mu-\hat{\mu})^2\right\}\right]\right)\right.\\
&&\left. + E\left(\left\{\Pi(D)-\hat{\Pi}(D)\right\}\left[{\Pi}\left\{g({\mu})\right\}-\hat{\Pi}\left\{g({\mu})\right\}+\hat{\Pi}\left\{g({\mu})-g(\hat{\mu})\right\}\right]\right)\right.\\
&&\left.+\left(1-\frac{E\left[\left\{D-\Pi(D)+\Pi(D)-\hat{\Pi}(D)\right\}^2\right]}{{E}\left[\left\{D-{\Pi}(D)\right\}^2\right]}\right)E\left(\left\{D-\Pi(D)\right\}\left[g(\mu)-\Pi \left\{g(\mu)\right\}\right]\right)\right\}\\
&&=\frac{\sqrt{n}}{\hat{E}\left[\left\{D-\hat{\Pi}(D)\right\}^2\right]}\left\{
E\left(\left\{D-\hat{\Pi}(D)\right\}O_p\left\{(\mu-\hat{\mu})^2\right\}\right)\right.\\
&&\left. + E\left(\left\{\Pi(D)-\hat{\Pi}(D)\right\}\left[{\Pi}\left\{g({\mu})\right\}-\hat{\Pi}\left\{g({\mu})\right\}+\hat{\Pi}\left\{g({\mu})-g(\hat{\mu})\right\}\right]\right)\right.\\
&&\left.+\left(\frac{-E\left[2\left\{D-\Pi(D)\right\}\left\{\Pi(D)-\hat{\Pi}(D)\right\}+\left\{\Pi(D)-\hat{\Pi}(D)\right\}^2\right]}{{E}\left[\left\{D-{\Pi}(D)\right\}^2\right]}\right)\right.\\&&\left.\times E\left(\left\{D-\Pi(D)\right\}\left[g(\mu)-\Pi \left\{g(\mu)\right\}\right]\right)\right\}\\
&&=\frac{\sqrt{n}}{\hat{E}\left[\left\{D-\hat{\Pi}(D)\right\}^2\right]}\left\{
E\left(\left\{D-\hat{\Pi}(D)\right\}O_p\left\{(\mu-\hat{\mu})^2\right\}\right)\right.\\
&&\left. + E\left(\left\{\Pi(D)-\hat{\Pi}(D)\right\}\left[{\Pi}\left\{g({\mu})\right\}-\hat{\Pi}\left\{g({\mu})\right\}+\hat{\Pi}\left\{g({\mu})-g(\hat{\mu})\right\}\right]\right)\right.\\
&&\left.+\frac{E\left[\left\{\Pi(D)-\hat{\Pi}(D)\right\}^2\right]}{{E}\left[\left\{D-{\Pi}(D)\right\}^2\right]}E\left(\left\{D-\Pi(D)\right\}\left[g(\mu)-\Pi \left\{g(\mu)\right\}\right]\right)\right\},
\end{eqnarray*}
where we use that $g(\mu)-\Pi \left\{g(\mu)\right\}$ is orthogonal to $\Lambda$ and thus has covariance zero with $\Pi(D)-\hat{\Pi}(D)$, and likewise
$D-\Pi(D)$ does not covary with $\Pi \left\{g(\mu)\right\}-\hat{\Pi}\left\{g(\hat{\mu})\right\}$.
It now follows 
that the remainder term converges to zero in probability under the conditions 
\begin{eqnarray*}
E\left(\left\{D-\hat{\Pi}(D)\right\}O_p\left\{(\mu-\hat{\mu})^2\right\}\right)&=&o_p(n^{-1/2})\\
E\left(\left\{\Pi(D)-\hat{\Pi}(D)\right\}^2\right)^{1/2}E\left(\left[{\Pi}\left\{g({\mu})\right\}-\hat{\Pi}\left\{g({\mu})\right\}\right]^2\right)^{1/2}&=&o_p(n^{-1/2})\\
E\left(\left\{\Pi(D)-\hat{\Pi}(D)\right\}^2\right)^{1/2}E\left(\left[\hat{\Pi}\left\{g({\mu})-g(\hat{\mu})\right\}\right]^2\right)^{1/2}&=&o_p(n^{-1/2})\\
E\left[\left\{\Pi(D)-\hat{\Pi}(D)\right\}^2\right]&=&o_p(n^{-1/2}),
\end{eqnarray*}
where the first condition is redundant when $g(.)$ is the identity link.

\end{document}